\documentclass[%
 reprint,
 amsmath,amssymb,
 aps,
]{revtex4-2}

\usepackage{amsmath}
\usepackage{amsfonts}
\usepackage{graphicx}
\usepackage{color}
\usepackage[usenames,dvipsnames]{xcolor}
\usepackage{float}
\usepackage{soul}
\usepackage{cancel}
\usepackage{hyperref}
\usepackage[normalem]{ulem}
\hypersetup{
   colorlinks,
   citecolor=blue,
   filecolor=blue,
   linkcolor=blue,
   urlcolor=blue,
   breaklinks=true
}
\usepackage{dsfont}
\usepackage{amssymb}

\usepackage{subfigure}
\newcommand{\lh}[1]{{\color{purple} [LH: #1]}}

\newcommand{\lhreplace}[1]{{\color{black} #1}}
\newcommand{\lhremove}[1]{}

\newcommand{\freplace}[1]{{\color{black} #1}}
\newcommand{\fremove}[1]{}

\begin{document}
\title {{\color{black}A Model for Topological p-wave Superconducting Wires with Disorder and Interactions}}

\author{Frederick del Pozo}
\affiliation{CPHT, CNRS,  École Polytechnique, Institut Polytechnique de Paris, 91120 Palaiseau, France}
\author{Loïc Herviou}
\affiliation{Univ. Grenoble Alpes, CNRS, LPMMC, 38000 Grenoble, France}
\affiliation{Institute of Physics,  École Polytechnique Fédérale de Lausanne (EPFL), CH-105 Lausanne, Switzerland}
\author{Olesia Dmytruk}
\affiliation{CPHT, CNRS,  École Polytechnique, Institut Polytechnique de Paris, 91120 Palaiseau, France}
\author{Karyn Le Hur}
\affiliation{CPHT, CNRS,  École Polytechnique, Institut Polytechnique de Paris, 91120 Palaiseau, France}

\date{\today}

\begin{abstract}
We present a comprehensive theoretical study of interacting and disordered topological phases of coupled Kitaev wires, which may support further realistic applications of Majorana fermions. We develop a variety of analytical, mathematical and numerical methods for one and two-coupled wires, associated with a topological marker accessible from real-space correlation functions on the wire(s). We verify the stability of the topological superconducting phase and quantify disorder effects close to the quantum phase transitions, e.g. through two-point correlation functions or using a renormalization group (RG) analysis of disorder. We show for the first time that the double critical Ising (DCI) phase --- a fractional Majorana liquid characterized by a pair of half central charges and topological numbers --- is stabilized by strong interactions against disorder which respects the inversion symmetry between the wires (ie. parity conservation on each wire).
In the presence of an inter-wire hopping term, the DCI phase turns into a protected topological phase with a bulk gap. We study the localization physics developing along the critical line for weaker interactions.
\end{abstract}


\maketitle

{\color{blue}\section{Introduction}} 
Overcoming quantum error propagation and decoherence \cite{schlosshauer2019quantumtoclassical} represents an important challenge in quantum information science, in particular with respect to quantum computation \cite{TopoQuantCompReview}.
{\it Topological} states, which are manifestly resilient against local perturbations, are promising candidates to realise platforms for practical applications at the hardware level. These states occur in phases characterized by an \emph{invariant}, which through the bulk-edge correspondence ensures the existence of localized quasi-particles at the edges of the system \cite{topo_review,alicea2012new}. One promising approach for realising such a topological platform is based on arrays of two or more \emph{Kitaev} wires {\color{black}\cite{Mazza_QMem1, Mazza_QMem2, TopoQuantCompReview}} --- a prototypical example of a $p$-wave superconducting (SC) wire hosting Majorana fermions (MFs) at its edges \cite{kitaev2001unpaired}. MFs can also occur in SC wires with different symmetries e.g. due to the presence of a magnetic impurity \cite{KLH2000}. Related models are the Quantum Ising chain --- with recent experimental signatures of edge states reported in \cite{QuantumIsingExp2022} --- and coupled quantum dots or spins \cite{Delft,MajoranaTopological2024}. 

Coupling Kitaev wires leads to interesting new phases, as seen for example in quantum ladders \cite{Ledermann_2000, ladder1, ladder2}, or in an array of wires coupled to a magnetic flux, which was shown to host $p+ip$ superconductivity \cite{Yang_2020}. If and how these phases may become relevant in future applications also relies on a robust understanding of the phase diagram. Similarly, the landscape of interacting topological phases remains largely unexplored, as the classification of topological phases based on dimension and symmetries \cite{Altland_1997} does not apply universally in interacting systems \cite{Fidkowski_2010}. In SC nano-wires and Kitaev wires, interaction effects were shown to reinforce the topological phase supporting MFs \cite{Miao_2018, Alicea_interacting, Gergs2016method, Schuricht}, whilst coupling to light has been proposed to reveal the topology of the wires \cite{Dmytruk15, Dmytruk16, Takis, Kozin, delpozo2024topological}. Finally, our attention must also turn to disorder, as in realistic systems impurities and fluctuations of system parameters are present. The interplay of interactions and disorder may then offer new insights into the properties of more realistic materials \cite{Gergs2016method, Laflorencie1, Laflorencie2}.

The topological invariant is often defined in a continuous Brillouin zone. This leads to difficulties when investigating finite-size and disordered systems, which has resulted in a multitude of different approaches \cite{PanSarma2021, Gergs2016method, decker2024density, prodan2010entanglement,bianco2011mapping, zhang2013coupling, Laflorencie1, Laflorencie2}. The occurrence of interfaces and (random) disorder generally affect the periodicity of the lattice. This has fueled the search for additional tools to classify topological phases using real-space techniques, eg. in one-dimensional wires \cite{Gergs2016method, del2023fractional, decker2024density, PanSarma2021}. Applied to interacting Kitaev wires, the real-space marker in \cite{del2023fractional} also revealed a fractional topological marker of a doubly critical Ising (DCI) phase far-from half-filling \cite{herviou2016phase, del2023fractional}. The DCI phase can be viewed in real space as a Majorana fermion liquid with two zero-energy Majorana fermion modes delocalized along the wires \cite{LeHur_maj}. We emphasize that this phase may be engineered with two quantum Ising chains coupled through a spin interaction along $z$ direction \cite{QuantumIsingExp2022}. 
A correspondence between the wires and two coupled Bloch spheres links the DCI phase to a free Majorana fermion at one pole \cite{KLHOneHalf,hur2022topological}. \lhremove{\color{black}  In the model of spheres, the stability and 
protection of Majorana fermions \cite{KLHOneHalf} towards disorder effects within the one-half fractional topological phase \cite{MajoranaTopological2024}, is shown as a result of transverse spin interactions \cite{HutchinsonKLH_2021}.}
 \lhreplace{On the spheres, the Majorana fermions \cite{KLHOneHalf} within the one-half fractional topological phase \cite{MajoranaTopological2024} are stabilized and protected from the disorder by transverse spin interactions \cite{HutchinsonKLH_2021}.}
Due to the close relationship between the DCI phase and the fractional phase of the Bloch-sphere model \cite{HutchinsonKLH_2021, del2023fractional}, we naturally wonder about the stability against disorder, taking into account that the transverse spin interaction does not have a simple analogue with fermions. 

In this work, we generally address the stability of topological p-wave superconducting wires in the presence of {\it disorder} and {\it interactions} by developing analytical and numerical approaches associated to a real-space marker that we recently introduced in \cite{del2023fractional}. We envision that our work, eg. through the tools that we are developing, will be useful for realistic applications with Majorana fermions \cite{QuantumIsingExp2022, Delft}. 

The paper is organized as follows. In Sec. \ref{Sec2} we recall the model for a Kitaev $p$-wave superconducting wire and show the effects of global (ie. across the whole wire) and local (ie. at each lattice site) disorder on the real-space topological markers introduced in \cite{del2023fractional}. By considering a stochastic Chern marker --- from an analogy to the Haldane model on the honeycomb lattice \cite{klein2021interacting} --- and the two-point correlation function describing the kinetic term on nearest neighbors, we justify the stability of the topological phase and to reveal the presence of a topological phase transition in the presence of disorder both analytically and through {\it exact diagonalization} (ED). In Sec. \ref{Sec3} and \ref{Sec4} we study the case of two Kitaev wires interacting via a Coulomb interaction between both legs or both wires \cite{herviou2016phase,del2023fractional}. We first demonstrate using DMRG the ability of the Chern marker to distinguish the topological and trivial phases in the presence of weak to moderate disorder, by comparing the marker to the lowest levels of the entanglement spectrum and Majorana edge mode amplitudes. Then, in Sec. \ref{Sec4}, we present analytical calculations and numerical simulations with the {\it Density Matrix Renormalization Group} (DMRG) algorithm in Julia (ITensors.jl) \cite{Fishman_2022}. We combine both a perturbative RG approach together with a numerical analysis of local correlation functions using the DMRG algorithm. Our analysis predicts, that at large interactions, the DCI phase is stabilized as long as fermion parity conservation in each wire is preserved, equivalent to an inversion symmetric coupling of the wires. We demonstrate that inversion symmetry breaking disorder is closely related to inter-wire hopping processes, which were shown in \cite{del2023fractional} to introduce a localization length $L^{*}$ for the critical modes. A topological ($2$MF) phase with shared edge modes emerges, and our analysis reveals the stability of the $2$MF phase against disorder. We therefore conclude that for lengths far below the localization scale $L^{*}$ the coupled wires resemble the DCI phase. 
Finally in Sec. \ref{Sec5} we present concluding remarks, whilst appendices provide supplementary information on the derivations of correlation functions and RG equations, and details on the (numerical) DMRG method.\\

{\color{blue}\section{New Hallmarks of Topology in a Single Kitaev wire}\label{Sec2}}

{\color{black}  We investigate the effects of disorder on a single Kitaev wire with $N$ sites \cite{kitaev2001unpaired}, in the presence of a screened Coulomb interaction between neighbouring sites.\lhremove{assuming spin-polarized or spinless fermions.}}
The phase diagram of this model in the clean limit has already been investigated in great detail, eg. in \cite{Schuricht}. 
Verifying the stability of topological properties is extremely relevant for applications, and requires a robust method to ascertain the topological indices also in the presence of disorder. In two-dimensional systems, the Chern number with disorder can be evaluated using its real-space representation~\cite{prodan2010entanglement,bianco2011mapping} or a coupling-matrix method~\cite{zhang2013coupling}. 
Conversely in $1$D, the work in \cite{PanSarma2021} shows how methods based on the transfer-matrix can be used for non-interacting $p-$wave superconductors.
However, these approaches in disordered systems yet present numerical challenges, especially in the presence of interactions. Experimentally, they may also be difficult to implement. The interplay of disorder and Hubbard-like or Coulomb interactions has also attracted interest e.g. in \cite{Gergs2016method, decker2024density}, which introduced a topological invariant expressed in terms of real-space correlation functions in the Jordan-Wigner transformed model, ie. a 
highly non-local expression on the wire. 

In this work, we apply the recently proposed real-space formulation of the Chern number for the Kitaev wire \cite{del2023fractional}.
We first show that the topological number $C$ can be used equivalently to map the topological phase diagram in the presence of disorder. We then present an approach to analytically include the interaction at a mean-field level \cite{klein2021interacting,hur2022topological}, which offers further insights into the quantitative evolution of the topological marker versus global Gaussian disorder with strength $\sigma/t$. 
The topological markers are also extracted and used to showcase the difference between local and global disorder. Local correlation functions obtained from exact diagonalization (ED) give further insights into the properties of the topological superconducting wire across the phase transition.

The Kitaev wire \cite{kitaev2001unpaired} is defined by the Hamiltonian
\begin{align}
	H_K &= -\sum_{j=1}^{N}\mu_j c_j^\dag c_j - t\sum_{j=1}^{N-1}\left(c_j^\dag c_{j+1} + \text{h.c.}\right)\notag\\ &+\Delta\sum_{j=1}^{N-1}\left(c^\dag_j c^\dag_{j+1} + \text{h.c.}\right) + V\sum_{j=1}^{N-1} c_j^\dag c_j c_{j+1}^\dag c_{j+1}.
	\label{eq:KitaevH}
\end{align}

Here, $c_j$ ($c_j^\dag$) are fermionic annihilation (creation) operators at site $j$,   $\mu_j$ is the site-dependent chemical potential, $t$ is the hopping amplitude, and $\Delta$ is a $p$-wave superconducting pairing potential. The  Kitaev wire  in the absence of disorder ($\mu_j \equiv \mu$) presents a topological phase with one Majorana fermion appearing at each end of the wire if $|\mu|< 2t$. We note that the superconducting pairing $\Delta$ does not enter into the topological criterion for the Kitaev wire. At $\mu = \pm 2t$ the Kitaev wire is at the critical point. 

In this work, we apply the recently proposed real-space formulation of the ``Chern number" for the Kitaev wire \cite{del2023fractional}, which is expressed exclusively in terms of two-point correlation functions on the wire itself
\begin{equation}\label{eq:Cnumber}
	C = \dfrac{1}{N}\sum_{i = 1}^{N/2}\sum_{j = 1}^{N}\langle  c^\dag_{j} c_{j+2i-1} +  c^\dag_{j+2i-1} c_j \rangle.
\end{equation}
\freplace{This definition is manifestly periodic, and can also be applied to the bulk of open chains by identifying $c_{r+N} \equiv c_{r}$ for $1\leq r\leq N$. 
In the non-interacting and clean limit, the correlation functions in \eqref{eq:Cnumber} above can be calculated analytically.}
{\color{black}  From the Bogoliubov de Gennes (BdG) formalism detailed in Appendix \ref{AppendixA}, we obtain \fremove{the formula}
\begin{equation}\label{eq:correls_m}
\langle c^{\dagger}_{j+m} c_j\rangle = \frac{1}{\pi}\int_0^{\pi} dk\cos(k m)|v_k|^2 \text{ for } \vert m \vert  \ll N.
\end{equation}
To define $v_k$, we introduce the angle $\theta_k$ defined by
\begin{align}
\label{mathequations}
\cos \theta_k &= \frac{- \mu - 2 t_\mathrm{eff}\cos k}{E_k}, \phantom{...}
\sin \theta_k = \frac{2 \Delta_\mathrm{eff}\sin k}{E_k}\\
E_k &= \sqrt{ (\mu + 2 t_\mathrm{eff}\cos k)^2  + 4 \Delta_\mathrm{eff}^2 \sin^2 k.}
\end{align}
Within these definitions, we also account for interaction effects and introduce  $t_\mathrm{eff} = t+ V\langle c_{j+1}^\dag c_{j}\rangle$ and  $\Delta_\mathrm{eff} = \Delta + V \langle c_{j+1}c_j\rangle$. Within a mean-field theory approach at weak interactions, the location of the quantum critical point (QCP) is modified to $\mu=2t+\frac{V}{2}$. The linear evolution of the location of the QCP with the interaction agrees with the results of Ref. \cite{Schuricht}. Then, we define the BdG functions $u_k = \cos \frac{\theta_k}{2}$ and $ 
v_k = i \sin \frac{\theta_k}{2}$, which satisfy $|u_{k}|^{2} + |v_{k}|^{2} = 1$ as expected. 
In the following paragraph, for simplicity sake, we take $V=0$ and results can be easily adapted modulo the change $t\rightarrow t_{\mathrm{eff}}$ and $\Delta_\mathrm{eff} \rightarrow \Delta + V \langle c_{j+1}c_j\rangle$.

A priori, all correlators are necessary to evaluate Eq.~\eqref{eq:Cnumber}, reflecting the topological (global) property of the $C$ number.
}
\lhreplace{However, in gapped phases, the series will be dominated by the short-range contribution due to the exponential decay of correlations. 
In fact, for $\mu=0$ and $t=\Delta$, we can rigorously show that $\langle c^{\dagger}_{j+m} c_j\rangle = 0$ if $|m|>1$.
The topological invariant in Eq. (\ref{eq:Cnumber}) then precisely becomes $C=4\langle c^{\dagger}_{j+1} c_j\rangle=1$ with the correlator itself being related to the winding number.
Note that this relation only stands at special points: in the limit $\frac{\Delta}{t}\rightarrow 0$ for example, $\langle c^{\dagger}_{j+1} c_j\rangle=\frac{1}{\pi}$. 
Generally, we expect an exponential convergence of $C$ with the range of the evaluated correlators.}
{\color{black}  At the non-interacting quantum critical point (QCP), it is also possible to evaluate all correlators analytically. 
As we show in Appendix \ref{AppendixA}, we obtain for the ideal (clean) wire, $\Delta=t$ and $\mu = -2t$:
\begin{eqnarray}
\langle c^{\dagger}_j c_{j+m} \rangle &=& -\frac{1}{2\pi}\int_0^{\pi} dk \cos(km)\sin\frac{k}{2} \nonumber \\
&=& \frac{1}{\pi}\frac{1}{4m^2 - 1}.
\end{eqnarray}
}
The continuous version of Eq.~\eqref{eq:Cnumber} becomes
\begin{equation}
\label{C2}
C = \sum_{i=-\infty}^{+\infty} \langle c^{\dagger}_j c_{j+ 2i + 1} + 
c_{j + 2i + 1}^{\dagger} c_j \rangle,
\end{equation}
{\color{black}where an implicit sum $\frac{1}{N}\sum_{j = 1}^{N}$ and a subsequent thermodynamic limit have been performed in Eq. (\ref{eq:Cnumber}). 
The terms with negative values of $i$ in the sum are another equivalent way to account for the (additional) terms coming from the PBC limit in Eq. (\ref{eq:Cnumber}).}
This gives precisely $C = \frac{1}{2}$ at the transition, though only converge as $m^{-1}$ where $m$ is the largest distance considered.
This half-Chern number can be understood has a half-Skyrmion in the Bloch sphere formulation of the BCS model \cite{hur2022topological,del2023fractional}; see Appendix \ref{AppendixA}.

\freplace{A central aim of our work is to study the stability of the Chern marker in Eq.~\eqref{eq:Cnumber} in the presence of local fluctuations of the chemical potential. 
As such, another important tool we use to study the stability of the topological and physical properties of individual and coupled Kitaev wires are precisely the correlation functions in Eq.~\eqref{eq:correls_m}, and in particular the nearest-neighbor correlator. 
For an illustration, see formula (\ref{formulatDelta}) and Fig. \ref{fig:CorrelationFunctionDeltaNeqT}. 
Later on in Sec. \ref{Sec4} we also examine the same correlation function $\langle c^{\dagger}_{j+1}c_{j}\rangle$ across the phase boundary in the presence of disorder for two interacting wires.} \\

\begin{figure}[h!] 
	\includegraphics[width=0.99\linewidth]{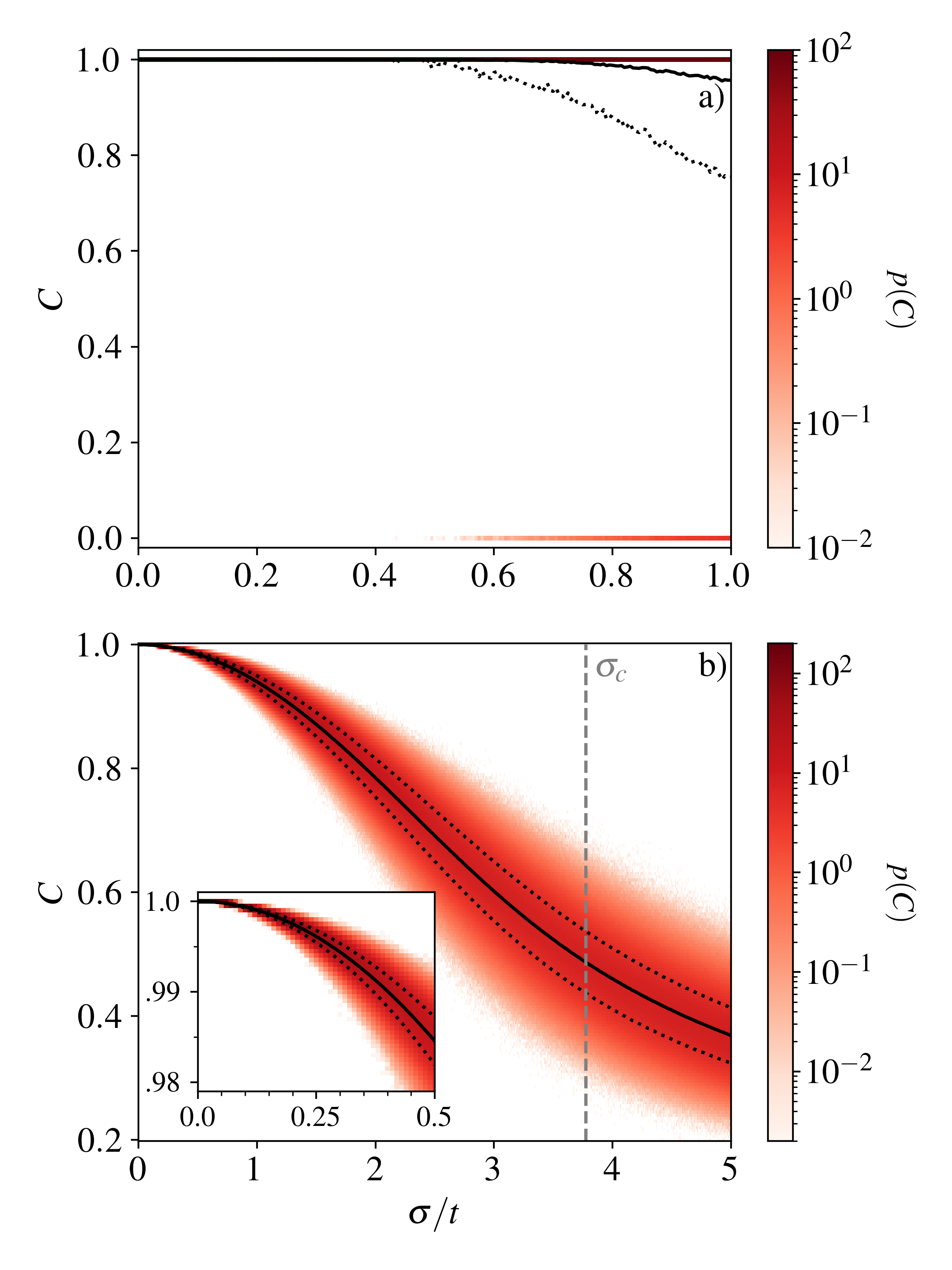}
	\caption{Topological number $C$ as a function of the variance $\sigma$ for the Kitaev wire with PBCs in presence of disorder in the chemical potential calculated using Eq. \eqref{eq:Cnumber}. Black solid line marks the mean value of $C$, while the dotted line are one standard deviation away. The red color is the estimated probability density $p(C)$. (a) We consider global disorder with $\delta\mu_j \equiv \delta\mu$ with $10^4$ samples. The mean value agrees exactly with the stochastic Chern number (blue dotted line) given by Eq.~\eqref{eq:Capp}.  (b) We consider on-site disorder $\delta\mu_j$ (Gaussian, with standard deviation $\sigma$), and $10^5$ samples for each $\sigma$. The Chern number $C$ decreases continuously. At $\sigma_c \approx 3.775$ (the dashed gray line), we have $C\sim 0.5$ indicating the transition from topological to disordered phase. In inset, a zoom on the region $\sigma \leq 0.5$.
    The parameters are chosen as $\mu = 0$,  $\Delta = t = 1$, and $N = 100$.}
	\label{fig:CnumberDisorderMu=0C}
\end{figure}

We start by considering  the effect of Gaussian disorder on the topological properties of the non-interacting Kitaev wire by adding on-site fluctuations $\delta \mu_j$ to the chemical potential as $\mu_j = \mu +\delta \mu_j$ in Eq.~\eqref{eq:KitaevH}. 
On-site fluctuations $\delta\mu_j$ are sampled from a normal distribution with zero mean and Gaussian standard deviation $\sigma$ 
\begin{equation}
	P(\delta\mu) = \dfrac{1}{\sqrt{2\pi}\sigma}e^{-\frac{(\delta\mu)^2}{2\sigma^2}},
	\label{eq:NormalDistribution}
\end{equation}
where $\sigma/t$ quantifies the disorder ``strength". The topological phase is in general resilient to disorder. {\color{black} In Appendix \ref{AppendixA}, in Fig. \ref{fig:CorrelationFunction2Panels}, we present the results from ED for one Gaussian disorder realization at $\mu=0$ for the correlator $\langle c^{\dagger}_j c_{j+1}\rangle$ showing its robustness for $\mu=0$ until $\sigma\sim 0.5t$. The (averaged) topological number as a function of $\sigma$ is then shown for {\it global} disorder in Fig. \ref{fig:CnumberDisorderMu=0C}a) corresponding to $\delta\mu_j=\delta\mu$ with $10^4$ samples, and for {\it on-site} Gaussian disorder in Fig. \ref{fig:CnumberDisorderMu=0C}b) corresponding to $10^5$ samples.}
A simple transfer matrix approach, eg. for $t = \Delta$, shows that the edge Majorana fermions remain localized up to \cite{Gergs2016method}
\begin{equation}
    \int\limits  d \delta \mu P(\delta \mu) \log \frac{\vert \mu + \delta \mu \vert }{2} = \log \vert t \vert. \label{eq:criticalsigma}
\end{equation}
For a normally distributed $\delta \mu_{j}$ centered around zero, Eq.~\eqref{eq:criticalsigma} yields the critical $\sigma_c/t = e^{\frac{1}{2}(\gamma_e  + \log 8)}\approx 3.775$,
where $\gamma_e$ is the Euler's constant. The disorder averaged Chern number $C$ as a function of $\sigma$ is shown in Fig. \ref{fig:CnumberDisorderMu=0C}. For {\color{black} small to moderate values} of $\sigma$ the marker is $C \approx 1$, but decreases as $\sigma$ is increased. Remarkably, it crosses the transition line at $C \approx 0.5$, and further connects the fractional marker with the QCP. 
We note that the critical point $\mu = \pm 2t$ is unstable in the presence of the Gaussian disorder, since any amount of disorder drives the wire out of this point. 
We also show that the behaviour of local correlation functions for large values of $\sigma$ deep in the topological and close to the QCP suggest localization in the latter. 

To better understand the evolution of the topological marker $C$ in Fig.~\ref{fig:CnumberDisorderMu=0C}, we employ a stochastic approach as in \cite{klein2021interacting}. 
There, a duality between global disorder and the effects of ``heating" in the two-dimensional interacting Haldane model was found. The stochastic Chern marker is defined as the disorder averaged marker 
$
	C = \int_{-\infty}^{+\infty} d\delta\mu \ P(\delta\mu) C(\delta\mu)$,
where $P(\delta\mu) $ is given by Eq.~\eqref{eq:NormalDistribution}. 
We note that the fluctuations in the chemical potential within the interval $-2t-\mu < \delta\mu < 2t-\mu$ result in the Chern number $C(\delta\mu) = 1$, and $0$ otherwise. Therefore,  the stochastic Chern number reads
\begin{align}
	&C_{st} = 1 - \int_{-\infty}^{-2t-\mu} d\delta\mu \ P(\delta\mu) C(\delta\mu) \notag\\
 &-  \int_{2t-\mu}^{+\infty} d\delta\mu \ P(\delta\mu) C(\delta\mu). 
\end{align}
We introduce $\delta\mu_c = |2t-\mu|$ and assume that fluctuations $\delta\mu$ are small, leading to the approximate Chern number
\begin{align}
	C_{st} \approx 1 - 2 P( |2t-\mu|) C( |2t-\mu|).
	\label{eq:Capp}
\end{align}
This formula illustrates that the stochastic topological marker acquires a small exponential correction with disorder. For the two-dimensional interacting Haldane model ~\cite{klein2021interacting,hur2022topological}, this correction can be attributed to particle-hole pair excitations produced above the gap as a result of interaction effects, which can be viewed as a heating effect induced from interactions. 
For the topological $p$-wave superconductor, we verify for $V=0$ that within the Nambu basis matrix representation, the stochastic Chern number remains quantized until large $\sigma\sim 0.6t$ in agreement with ED, see the fit in dashed blue in Fig. \ref{fig:CnumberDisorderMu=0C} to the formula (\ref{eq:Capp}) obtained for global disorder $\delta\mu_j=\delta\mu$. The topological phase is stable under the effect of this disorder, as expected.

\begin{figure}[h!] 
\includegraphics[width=0.95\linewidth]{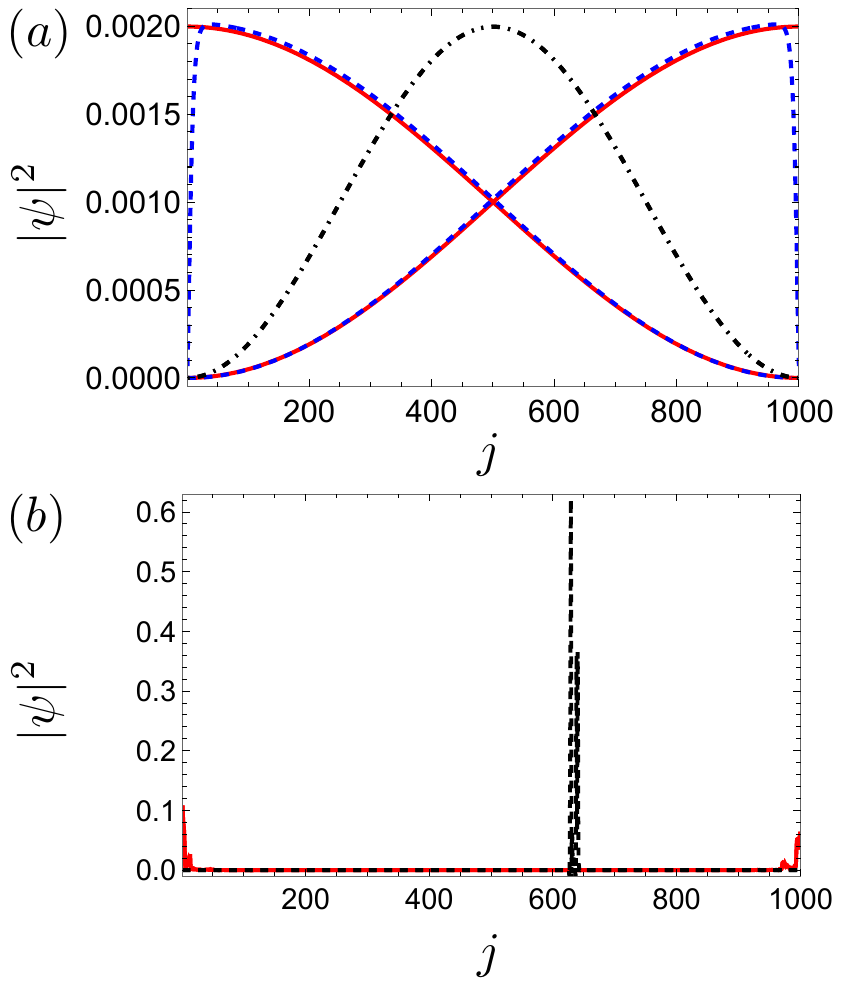}
\caption{Amplitude of zero-energy (or closest to) eigenstates $|\psi_{j}|^2 = \| \langle j|\psi\rangle\|^{2}$, for lattice site $j$. For $\mu/t \leq 2$, this corresponds to the localized Majorana modes. In (a) we show density $|\psi_{j}|^2$ of the clean finite-length Kitaev wire at the QCP $\mu = -2t$. Red solid line corresponds to $\Delta = t$, blue dashed line corresponds to $\Delta = 0.1 t$, and black dot-dashed line corresponds to $\Delta = 0$. In (b), for $t = \Delta$ and $\mu = -2t$ we show for single realisations of Gaussian disorder $\sigma/t = 0.5$ (red solid line) and $\sigma = 5t$ (black dashed line). The corresponding $C$ markers are $C = 0.89$ and $C = 0.35$ respectively. In both panels we normalized $|\psi_{j}|^2$, ie. $\sum_{j=1}^{N}|\psi_{j}|^2 = 1$. We fixed  $t = 1$ and $N = 1000$. The difference in amplitude for panels (a) and (b) : In panel (a) $|\psi|^2$ is non-zero on all lattice sites and  the probability density at one given site $j$ is small to give $1$ after summing over $1000$ lattice sites. In panel (b) $|\psi_{j}|^2$ is non-zero only on a few lattice sites.
}
 \label{fig:ProbabilityDensityCritical1000}
	\end{figure}

In Fig. \ref{fig:ProbabilityDensityCritical1000}~(a), for completeness, we also show the projection of the lowest-lying eigenstate onto the lattice site $j$, ie. $|\langle j|\psi\rangle|^{2}$. We find the existence of an extended mode for $\mu = -2t$ as long as 
$\Delta > 0$.
For $\mu=-2t$ and $k\rightarrow 0$, equivalently we obtain zero-energy states which can be written as superpositions of particle and hole \cite{del2023fractional}
\begin{equation}
\begin{aligned}
\label{Majoranafermions}
    \sqrt{2}\eta_k^- = c_k - ic^{\dagger}_{-k} &= \frac{1}{2}(\gamma_A-\gamma_B)(1-i) \\
\sqrt{2}\eta_{-k}^+ = c_{-k} +ic_{k}^{\dagger} &=  \frac{1}{2}(\gamma_A+\gamma_B)(1+i).
\end{aligned}
\end{equation}
The $\eta$ quasi-particles are related to the \emph{chiral} Majorana fermions $\sim \gamma^{B} \pm \gamma^{A}$. To find such eigenstates, we assumed that $k\Delta\gg |\mu\pm 2t|$. When $k\rightarrow 0$, any amount of weak, local disorder will then favor eigenstates with occupied or empty states ($\langle c^{\dagger} c \rangle=1$ or $0$) at $k=0$. 
At $\mu=2t$, the averaged $C$ number then alternates randomly between $1$ and $0$ for one disorder realization for a finite sample from the Bloch sphere description \cite{del2023fractional}. In this sense, the disordered averaged number can be seen as a mixture of being in the topological or trivial phase. We verify this result from ED with PBCs and $N=1000$ sites where $C$ takes values close to 1 or 0 already for $\sigma=0.001t$. In comparison, for the clean case, we have $C=1/2$ which is verified from the correlation $\langle c^{\dagger}_{j+1} c_j\rangle=\frac{1}{3\pi}$. The correlation function $\langle c^{\dagger}_{j+1}c_j\rangle$ also deviates more rapidly from the clean case limit with $\sigma$ as shown in Fig. \ref{fig:CorrelationFunction2Panels}(b).

In Fig. \ref{fig:ProbabilityDensityCritical1000}~(b) we further show the extracted GS probability at the QCP in the presence of disorder. We see that the extended modes in the clean case (in red) are now localized somewhere inside the wire for large disorder strength (in black). This corresponds to a trivial phase when averaging over disorder realisations.\\

{\color{blue}\section{Topological markers for two disordered and interacting wires}\label{Sec3}} 

Whilst numerous works have already investigated the interplay between disorder and interactions within an individual Kitaev wire \cite{PanSarma2021, Gergs2016method, decker2024density, Laflorencie1}, coupled chains remain largely unexplored in this respect. From a practical side, two or more wires have been proposed as topological qubit platforms \cite{TopoQuantCompReview}, or as topological quantum memory platforms \cite{alicea2012new, TopoQuantMem}. 
Due to the relatively close proximity of wires in realistic setups, it is natural to consider effects of interactions between two neighbouring chains. Due to the breakdown \cite{Fidkowski_2010} of the complete classification of topological phases based on symmetry and dimension \cite{Altland_1997}, investigating interaction effects in topological materials remains a timely and relevant effort. Beyond their topological properties, superconducting wires could also become building blocks for new, small scale circuitry. Understanding the phase diagram at large interactions and far-from half-filling is then also important to understand their potential for future applications.

We first introduce the relevant extension to the single Kitaev wire as studied in \cite{herviou2016phase} and \cite{del2023fractional}. Then, we establish the Chern marker as a useful and appropriate tool to characterize the phases up to moderate interactions using DMRG from the library \href{https://github.com/ITensor/ITensors.jl}{ITensors.jl} to include interactions and disorder. 
The critical regions of the phase diagram are discussed separately in Section \ref{Sec4}.\\

{\color{blue} \subsubsection{Reviewing the phases of two interacting Kitaev wires}}

\begin{figure}[h!]
	\centering
	\includegraphics[width = \linewidth]{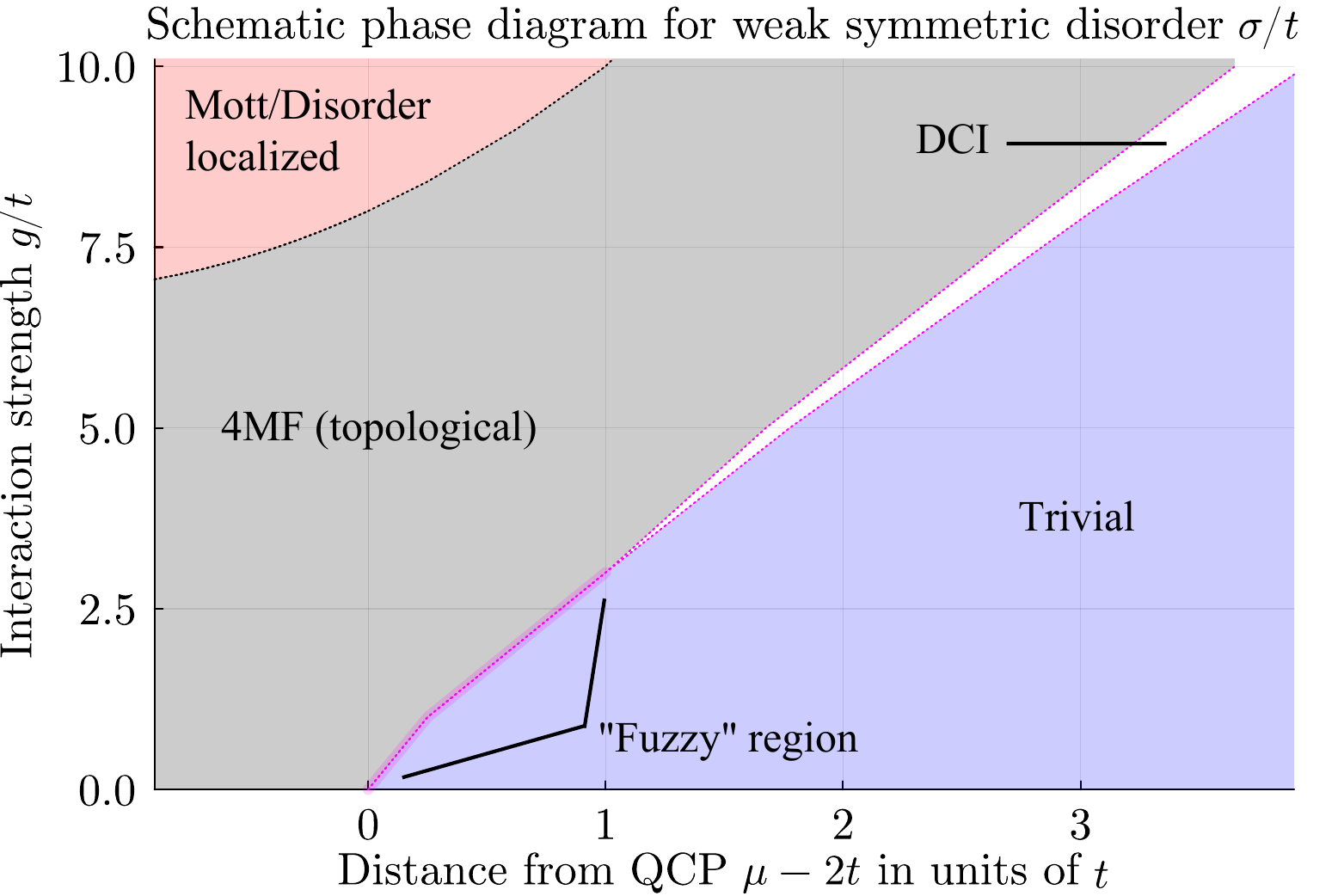}
	\caption{Schematic phase diagram for two interacting Kitaev wires in the presence of weak (symmetry preserving) Gaussian disorder with variance $\sigma^{2}$. It is a result of our combined analysis of theoretical (RG) and numerical (DMRG) methods, with precise phase boundaries dependent on system size and parameters. We highlight the region far-from half-filling around the DCI phase, containing also the doubly topological ($4$MF) and trivial phases. 
 Comparison to the clean diagram \cite{herviou2016phase} shows a ``fuzzy'' region along the critical line.}
	\label{fig:PhaseDiagramlowdis}
\end{figure}

In the clean situation, repulsive Coulomb interactions were found to enhance the topological phase \cite{herviou2016phase, del2023fractional}. They were explicitly modeled to lowest-order by an effective Coulomb strength $g > 0$, coupling nearest-neighbours on different chains:
\begin{equation}\label{eq:Hint}
    H_{int} = g\sum_{i} \left(n^{1}_{i} -\frac{1}{2}\right)\left(n^{2}_{i} - \frac{1}{2}\right).
\end{equation}
As illustrated in Fig. \ref{fig:PhaseDiagramlowdis}, the resulting phase diagram \cite{herviou2016phase} consists of a central, doubly-topological region ($4$MF) connected smoothly to $\mu/t = 0$, where each wire hosts one edge mode per side for open boundary conditions (OBCs) \cite{herviou2016phase}. On either side along the $\mu/t$ axis also a doubly trivial phase extends to large $g/t$. Close to half-filling, there is also a Mott insulating (MI) transition from the $4$MF phase for $g/t \gg 1$. This transition can no longer be described accurately by a sharp jump in $C$, which shows a smooth decrease $C \longrightarrow 0$ for $g\longrightarrow \infty$, as expected for a trivial phase \cite{del2023fractional}.
{\color{black}  We emphasize here that the formula in Eq. (\ref{eq:Cnumber}) for $C$ comes from an analogy with a model of coupled spheres which is rigorous at weak to moderate interactions \cite{HutchinsonKLH_2021,del2023fractional,hur2022topological}. 
Remarkably, the topological marker remains qualitatively correct at very strong interactions, e.g. close to the Mott phase or $\frac{g}{t}\sim 8$. 
In the presence of both interactions and disorder, disorder-averaging is therefore especially important to interpret the topological features of the different phases.
}
\lhremove{
Generalizing the definition of this topological marker to very strong interactions e.g. when entering in the Mott phase is yet a very good indicator of the physics for $\frac{g}{t}\sim 8$, but this also requires to be careful with assertions e.g. showing error bars on the results. Below, we precisely present the errors bars on the results e.g. for the (disorder averaged) topological invariant and also for the disorder averaged correlators.}

Between the $4$MF and trivial regions a doubly critical phase (DCI) is found, first along a critical line and then at large interactions as an extended region \cite{herviou2016phase}. Using the real-space Chern marker, in Ref. \cite{del2023fractional} we have previously showed that the DCI phase is additionally characterized by $C \approx 1/2$ in the case of PBCs.  {\  In this region, the ground state (GS) is degenerate between the parity sectors $(P_{1},P_{2}) = (0,1)$ and $(1,0)$, where we introduce the binary notation $F = (-1)^{P}$ for the parity $F$. By invoking the inversion symmetry of the coupled wires, the resulting physical GS then necessarily lies in the equal superposition of both parity sectors
\begin{equation}
    |DCI\rangle_{PBC} = \frac{1}{\sqrt{2}}\left( |10\rangle + |01\rangle\right).
\end{equation}
From this we find $C = \frac{1}{2}$ for each wire.}
By adding a hopping process $\sim t_{\bot}$ between neighbouring sites of the wires we obtain the Kitaev ladder, with $t_{\perp}$ introducing a gap in the DCI phase. This results in a simply-topological ($2$MF) phase with width $\sim t_{\bot}$ \cite{del2023fractional}, and corresponds to the previously found additional topological phase separating the $4$MF and trivial phases \cite{ladder1, ladder2, Yang_2020}, also with fractional marker $C = \frac{1}{2}$ due to a GS in superposition of $C^{\sigma} = 1$ and $C^{\sigma} = 0$ in the wire basis. 

Disorder is included in the same way as in equation \eqref{eq:NormalDistribution} in Sec.  \ref{Sec2}, ie. as a normally distributed chemical potential $\mu_{j} \in \mathcal{N}\left(\bar{\mu}, \sigma^{2}\right)$ for each lattice site $j$, with standard deviation $\sigma$. In section \ref{Sec4} we differentiate in particular between inversion symmetry preserving, and symmetry breaking disorder, ie. where $\mu^{1}_{j}$ and $\mu^{2}_{j}$ are either equal or independent on both wires. However, the main focus of the paper lies on symmetric disorder. 
Our results of the next sections are summarized schematically in Fig.  \ref{fig:PhaseDiagramlowdis} --- largely reproducing the clean phase diagram \cite{herviou2016phase} --- for weak disorder of the order $\sigma/t \sim 0.3$. The notable difference is the ``fuzzy" region along the critical line, which reflects the localization of the gapless critical modes below some critical interaction strength $g_{c}/t$. The phase diagram \ref{fig:PhaseDiagramlowdis} is found by combining analytical (RG) and numerical (DMRG) results, for the various limits of strong interactions and large chemical potentials. \\

{\color{blue}\subsubsection{Topological marker with interactions and disorder} }


As known from previous works, the topological phase remains stable as long as disorder fluctuations do not result in a gap-closing point \cite{PanSarma2021, Gergs2016method, decker2024density}. Mapping the interacting Kitaev wire onto a Luttinger liquid (LL) through bosonization, the resulting renormalization group (RG) equations \cite{herviou2016phase} at weak interactions $g$ in Appendix \ref{disorderflow} show that the superconducting pairing term $\Delta$ flows to strong couplings before weak disorder does. This reinforces the topological phase away from the QCPs as interactions drive the Luttinger parameter to values $K >1$, stabilizing the $p$-wave SC gap. Therefore, we expect for weak disorder the property $\langle C\rangle > 0.5$ to again signal the topological phase as for the single wire \ref{fig:CnumberDisorderMu=0C}. 

\begin{figure}[h!]
	\centering
	\includegraphics[width = 0.9\linewidth]{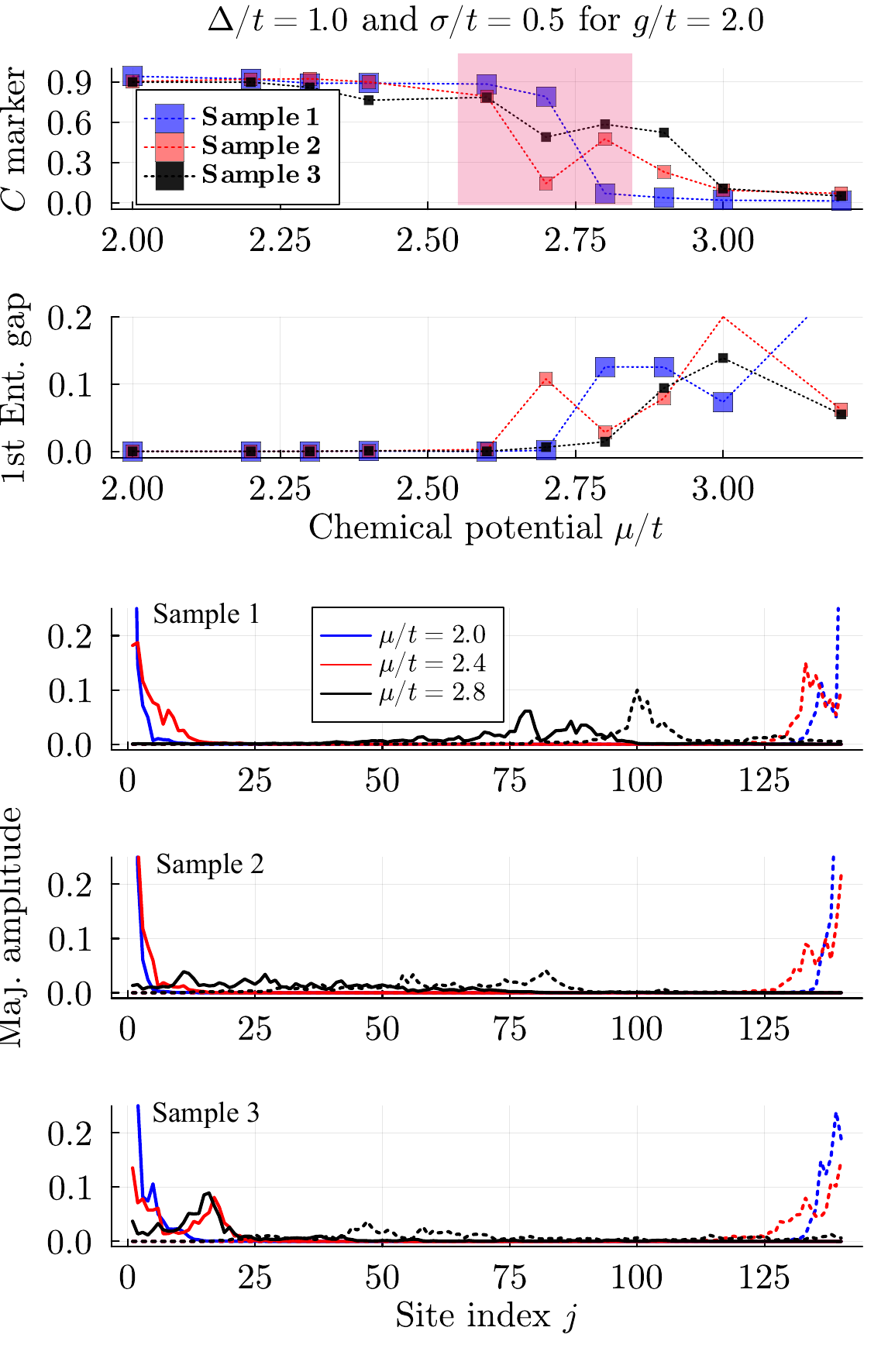}
\caption{Single realisations of disorder for $N = 140$, $t = \Delta$ and $g/t = 2.0$ and $\sigma/t = 0.5$. The shaded (pink) region in the first figure shows the ``fuzzy" region in \ref{fig:PhaseDiagramlowdis}. Outside of this region, the Chern marker, the first entanglement gap and the Majorana edge mode amplitudes clearly coincide. In the fuzzy region, we observe strong fluctuations of the $C$ marker and washed out Majorana amplitudes. In this region, finite-size effects and Majorana overlaps play a crucial role.}
	\label{fig:Delta1.0_sig05}
\end{figure}

By comparing $C$ in Eq. (\ref{eq:Cnumber}) for two wires to the lowest lying ($1$st) entanglement gap and amplitudes of Majorana fermions, we verify that it remains a meaningful marker for topology. The first entanglement gap is obtained from the entanglement spectrum \cite{Entanglementspectrum} and was found in \cite{Gergs2016method} to be a more reliable observable than the energy gap. For OBCs, the topological phase is characterized by a double degeneracy (or fourfold, for two wires) of the energy and entanglement spectra \cite{Entanglementspectrum, EntanglementPOV}. In Fig. \ref{fig:OBCvsDis_comp} we show a comparison for individual realisations of moderate disorder $\sigma/t = 0.5$, and interaction strength $g/t = 2.0$. The Chern marker, the entanglement spectrum and the edge modes agree very well for values on either side of the "fuzzy" region (shaded pink). Within this region, the $C$ number fluctuates strongly or is affected by finite-size effects. In finite systems, the edge modes are predicted to have a non-zero overlap resulting in a GS degeneracy lifting. {\color{black}In figures \ref{fig:OBCvsDis_comp} and \ref{fig:OBCvsDis_comp2} in Appendix \ref{AppendixC} we show a similar analysis at $\sigma/t = 0.3$ for $\Delta/t = 1.0$ and $\Delta/t = 2.0$ and $\Delta = 0.5$ respectively. The entanglement gap for $\Delta/t = 0.5$ deviates from zero much stronger than for $\Delta/t = 1.0$ and $\Delta/t = 2.0$, which matches the QFT predictions in Appendix \ref{AppendixB1}, highlighting the important competition between the effective SC gap $\Delta\left(l\right)$ and the disorder strength $\mathcal{D}\left(l\right)$.}
A scaling analysis for OBCs of the Chern marker in Appendix \ref{AppendixC} further reveals a strong dependency of $C$ on the entanglement spectrum --- in particular the lowest eigenvalue, see Fig. \ref{fig:OBC_Comp_N140} --- such that $C$ manifests itself also as an entanglement probe. 
\begin{figure}[h!]
	\centering
	\includegraphics[width = 1.0\linewidth]{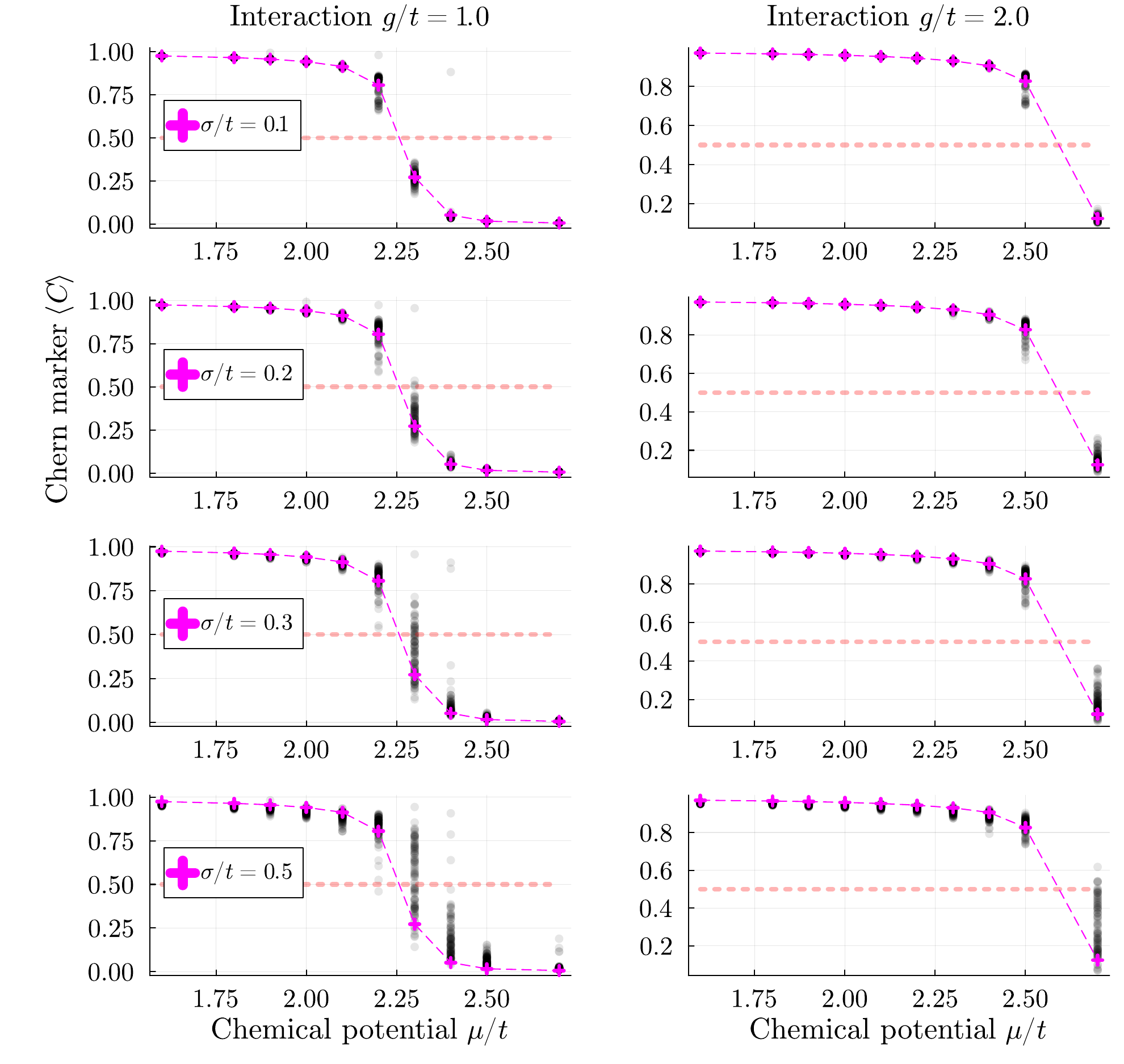}
\caption{Averaged $C$ marker for $M = 100$ realisations and $N = 140$ sites, for $t = \Delta = 1.0$ and OBCs.  (Black) Individual data points, (pink) the averaged $C$ value and (red) $\langle C \rangle = 0.5$. }
	\label{fig:avC}
\end{figure}

Whilst the edge mode amplitudes at $\mu/t = 2.4$ show a clearly localized mode at either boundaries of the wire, we also find in some cases signatures of the Majorana modes near the transition, similar as in \cite{Gergs2016method}. These are then dependent on specific disorder realisations and therefore vanish upon averaging over many samples. The same is the case for the $C$ marker, shown in Fig. \ref{fig:avC}, where the fluctuations of the fuzzy region average out to
the critical value $\langle C\rangle = 1/2$ at the transition. Whilst the individual realisations (black) show clear fluctuations in the fuzzy regime, the average values (pink) instead follow a clear transition from topological to trivial. Larger interactions stabilize the topological phase, which can be inferred from both the larger critical $\mu/t$, as well as the lower spread of the Chern marker close to the transition \cite{del2023fractional}. This might be particularly relevant in the context of superconducting circuits and other realistic applications, wherein the repulsive Coulomb interaction between electrons will play an important role due to the proximity of the individual components. 
{\color{blue}\subsubsection{Limitations of the real-space invariant}}
The results of previous sections therefore show that the Chern marker from real-space correlation functions in \eqref{eq:Cnumber} remains a powerful tool to investigate the topology of disordered, interacting Kitaev wires in the examined region of the phase diagram ($g/t \leq 2.0$ and $\sigma/t \leq 1.0$). Beyond these limits, we found the Chern marker to no longer capture the topological properties with sufficient accuracy. We find that the best results are achieved for PBCs, or for very large Kitaev chains  ($N\geq 200$) for OBCs.
For a more detailed discussion related to the differences between OBCs and PBCs, we refer to Appendix \ref{AppendixC}. 
Strong interactions also affect the quantization of the $C$ marker value considerably. Explanations for these deviations from quantized values of $C$ were discussed in more detail in \cite{del2023fractional}. Notably, $C$ does not have a discontinuity or steep-drop along the $g/t$ transition from topological to Mott order, schematically added in Fig. 
\ref{fig:PhaseDiagramlowdis}. Instead, for $\sigma/t = 0.0$ we found a smooth function of $g \longrightarrow \infty$ close to half-filling $\mu/t \approx 0$ \cite{del2023fractional}, with the transition again qualitatively at $C \approx 0.5$. This is reproduced for moderate disorder. In fact, from the Quantum Field Theory (QFT) description close to half-filling in Appendix \ref{disorderflow}, we find a similar model as in the Mott/disorder-localized transition studied in \cite{GiamarchiSchulz_loc1}. Here the trivial phase is determined by which of the two, disorder or Coulomb interactions, flows to strong coupling first. 
We therefore analyse this transition using perturbative RG equations as in \cite{herviou2016phase, GiamarchiSchulz_loc1}. 

{As we outline in Appendix \ref{disorderflow} and \ref{AppendixB1.2}, the two wires can be mapped onto a bosonic model with conjugate fields $\phi_{\pm}$ and $\theta_{\pm}$, and disorder can be introduced from first principles along the lines of \cite{GiamarchiSchulz_loc1}.
We assume both wires to be in relative close proximity, such that we may take the disorder contribution on each mode to be equal $\mathcal{D}_{\pm} = \mathcal{D}$. We then find disorder to contribute as the operator}
\begin{equation}
\begin{aligned}
 \sim \frac{\mathcal{D}}{\left(2\pi\right)^{2}\alpha^{3}}& \cos\left(\sqrt{2}\left(\phi_{+}\left(x,\tau'\right) - \phi_{+}\left(x, \tau\right)\right)\right)\\ \times&\cos\left(\sqrt{2}\left(\phi_{-}\left(x,\tau'\right) - \phi_{-}\left(x, \tau\right)\right)\right).
\end{aligned}
\end{equation}
Such a form of disorder was studied in \cite{GiamarchiSchulz_loc1}, where instead of the modes $\pm$ the two fields represented `spin' and charge sectors. Disorder renormalizes both the Luttinger parameter and the Fermi velocities $v_{F, \pm}$, however, due to the full set of flow equations summarised in the Appendix \eqref{Flows_half-filling}, we predict that the topological/Mott-insulating transition \cite{herviou2016phase} remains stable up to non-perturbative values of $\mathcal{D}$. Beyond this, our RG equations are no longer applicable, and a non-perturbative approach will be necessary to obtain the critical disorder strength. The exact transition is also dependent on the superconducting pairing strength, with disorder being more relevant for $\Delta/t \ll 1$, \emph{cf.} Appendix \ref{AppendixB1.2}.{\ We use a similar quantum field theoretical analysis to study the effects of disorder around the critical phase in Sec. \ref{Sec4}.}\\

{\color{blue}\section{Effects of disorder on the DCI phase far from half-filling}\label{Sec4}}
We now study the interplay between interaction and disorder effects in the critical region of the phase diagram. The DCI phase, separating the doubly topological and trivial phases of two interacting Kitaev wires in Fig. \ref{fig:PhaseDiagramlowdis}, was found in \cite{del2023fractional} to be protected by the discrete (exchange) $\mathbb{Z}_{2}$ symmetry between both wires. 
Therefore, we distinguish between symmetry preserving and breaking disorder, the latter of which assumes local fluctuations of $\mu$ on each wire independently. 

We numerically solve the RG equations derived in detail in Appendix \ref{AppendixB2} to gain insights on the mechanisms governing the competition between repulsive Coulomb interactions and local impurities. We predict a stabilization of the DCI phase against symmetry preserving disorder at large $g/t >0$. We also observed this for local correlation functions, reinforcing the understanding that the physics of the DCI phase can be understood in terms of the non-interacting QCP. \\

{\color{blue} \subsubsection{Reviewing the chiral description of the DCI phase}}

As was found in \cite{herviou2016phase}, in the DCI phase for two interacting wires the resulting Hamiltonian was shown to arise from bosonization of chiral fields $\psi_{R/L}$ made up from fermions of both wires. They were shown to be a direct extension of the single-wire chiral Majorana modes $\gamma_{R/L} = \gamma^{B} \mp \gamma^{A}$, which become gapless in the QCPs at $\mu = \pm 2t$, and give rise to the chiral Dirac fermions accross both wires
\begin{equation}\label{Chiral_modes_link_one_wire}
\begin{aligned}
           \sqrt{2}\psi_{R} = &\frac{1}{2}\left(\gamma^{2}_{B} -i \gamma^{1}_{B} + i\gamma^{1}_{A} - \gamma^{2}_{A}\right) = \frac{\gamma^{2}_{R} -i\gamma^{1}_{R}}{2}, \\
           \sqrt{2}\psi_{L} = & \frac{1}{2}\left(\gamma^{2}_{B} -i \gamma^{1}_{B} - i\gamma^{1}_{A} + \gamma^{2}_{A}\right) = \frac{\gamma^{2}_{L} -i\gamma^{1}_{L}}{2},
\end{aligned}
\end{equation}
revealing the \emph{double} critical Ising character of the phase. {\color{black}  We emphasize the role of the inversion symmetry between wires in the chiral fermion basis.
The inversion $1\leftrightarrow 2$ which maps $\psi_R\rightarrow -i\psi_R^{\dagger}$ and similarly $\psi_L\rightarrow -i\psi_L^{\dagger}$ corresponds to an effective particle-hole symmetry. \fremove{ (leaving e.g. $\psi^{\dagger} \psi$ invariant).}}
We further define the mixed-wire fermions
\begin{equation}
\begin{aligned}
    \Gamma_{j} =& \frac{1}{2}\left(\gamma^{A,1}_{j} + i\gamma^{A,2}_{j}\right) \\ \Theta_{j} =& \frac{1}{2}\left(\gamma^{B,2}_{j} - i\gamma^{B,1}_{j}\right),
\end{aligned}
\end{equation}
related to the chiral modes in equation \eqref{Chiral_modes_link_one_wire} through $\psi_{R/L, j} = \frac{1}{\sqrt{2}}\left( \Theta_{j} \pm i\Gamma_{j}\right)$.
In the vicinity of the DCI phase, the chiral field theory can be written equivalently in terms of bosonic degrees of freedom which is a hallmark of one-dimensional physics and bosonization \cite{giamarchi2004quantum, del2023fractional, herviou2016phase}
\begin{figure*}
	\centering 
    \includegraphics[width =0.85\textwidth]{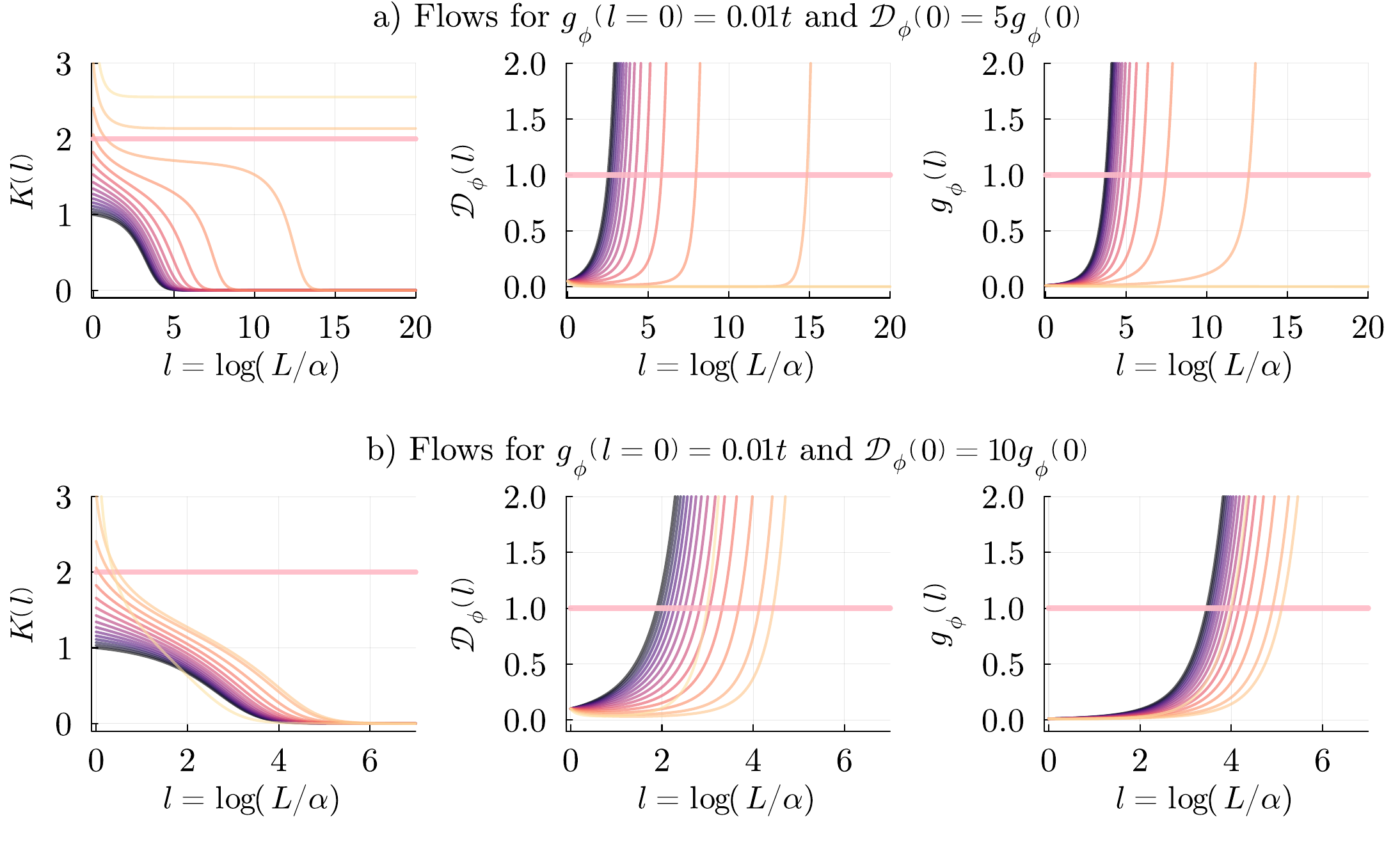}
	\caption{{\color{black} 
	Numerically solved flow equations along the critical line with $g_{\phi}\left(l = 0\right) = 0.01t$ and two disorder strengths $\mathcal{D}_{\phi}\left(0\right) = 5g_{\phi}(0)$ and $10g_{\phi}(0)$ respectively. Individual flows for $g/t \in \left[0.0, 3.0\right]$ with step-size $\delta g/t = 0.2$, and a color gradient \textit{dark} $\rightharpoonup$ \textit{light} indicating the progression of the flows with $g$. The $x-$axis denotes the exponential length-scale $l$ with $L = \alpha e^{l}$, and varies wildly depending on the input-parameters. The pink bands denote the $K = 2$ line beyond which we expect the DCI phase to be possible, or the $X(l) = 1.0$ line where the operator ``$X$" has flown to strong coupling. } }
	\label{fig:flow_symmetric}
\end{figure*}
\begin{equation}\label{Boson_Ham}
    H = \frac{v_{F}}{2\pi}\int \text{d}x \frac{1}{K}\left(\partial_{x}\phi\right)^{2} + K\left(\partial_{x}\theta\right)^{2} + g_{\phi}\cos\left(2\phi\right),
\end{equation}
where {\color{blue} 
$v_{F}K = 8t 
\alpha$, $K = \left(1 - \frac{8g}{\pi v_{F}}\right)^{-1/2}$} and $g_{\phi} = \left(\frac{2\delta \mu}{\pi \alpha}+\frac{2g}{\pi^2 \alpha}\right)$ with $\delta\mu = 2t \pm \mu$ measuring the deviation from the critical line. 
\lhreplace{The difference in bare parameters with Refs~\onlinecite{del2023fractional, herviou2016phase} is due to different conventions in the definition of the fields. It does not affect the qualitative final results, but simplify the inclusion of on-site disorder.}\\



{\color{blue}\subsubsection{Including disorder from first principles}}

To understand in more detail how disorder interacts with the critical Majorana modes in the DCI topological phase, we include disorder from first principles in the QFT description introduced in \cite{del2023fractional}. This reveals the intricate competition between disorder --- which seeks to localize the critical modes --- and interactions. For symmetric disorder across both wires, large enough interactions far away from half-filling can overcome the disorder and stabilize the DCI phase (see Appendix \ref{AppendixB2} for the derivation
of the RG equations). When the disorder acts asymmetrically onto the two wires, the resulting de-tuning of chemical potentials tends to break the $\mathbb{Z}_{2}$ symmetry between both wires and gaps the critical modes at arbitrary disorder strength. We emphasize that this is an important difference between the two wires situation and the quantum spin model of two spheres which can be protected against weak asymmetric disorder as a result of transverse spin interactions \cite{MajoranaTopological2024}.

We recall that symmetric disorder can be quantified with the symmetric disorder channel $2\mathcal{V}(x) = \delta\mu^{1}(x) + \delta\mu^{2}(x)$ with the upperscript $i$ corresponding to the wire index and $x$ to the position variable. Conversely, the asymmetric disorder parameter $2\nu(x) = \delta\mu^{1}(x) - \delta\mu^{2}(x)$ quantifies the relative deviations between the same sites on different wires. Disorder couples to the chemical potential, such that for the symmetric contribution this results in the additional term in the Hamiltonian
\begin{equation}\label{disorder_0}
    H_{dis} = \int_{x} \mathcal{V}\left(x\right)n_{x} \text{d} x = \int_{x} \mathcal{V}\left(x\right) \Gamma^{\dagger}_{x}\Theta_{x} \text{d} x  + \text{hc.}.
\end{equation}
Similarly, the asymmetric disorder $\nu\left(x\right)$ enters as a term proportional to $\Delta n_{x} = n^{1}_{x} - n^{2}_{x}$, ie.
\begin{equation}
   H_{dis, \nu} = \int_{x} \nu\left(x\right) \Delta n_{x}\text{d}x = \int_{x}\Gamma^{\dagger}_{x}\Theta^{\dagger}_{x} \text{d}x+ \text{hc.} 
\end{equation}
As demonstrated in Appendix \ref{AppendixB2}, these contributions to the Hamiltonian result in the following bosonic operators
\begin{equation}\label{disorder_pert}
\begin{aligned}
      H_{dis}  = &\int \text{d}x \ \frac{\mathcal{V}\left(x\right)}{2\pi\alpha}\cos\left(2\phi\left(x,t\right)\right) +\frac{\nu\left(x\right)}{2\pi\alpha}\cos\left(2\theta\left(x,t\right)\right).
\end{aligned}
\end{equation}

The Gaussian covariance matrices quantifying the disorder strength are defined with the convention
\begin{equation}
\begin{aligned}
       \langle \mathcal{V}\left(x\right)\mathcal{V}\left(y\right)\rangle = & \ 4\pi^{2}\mathcal{D}_{\phi}\delta\left(x- y\right)  \\ 
      \langle \nu\left(x\right)\nu\left(y\right)\rangle =& \   4\pi^{2}\mathcal{D}_{\theta}\delta\left(x-y\right).
\end{aligned}
\end{equation}
By performing the disorder average with the replica-field method  \cite{GiamarchiSchulz_loc1,senechal2006theoretical,giamarchi2004quantum}
\freplace{we derive in Appendix \ref{AppendixB2} the flow equations for the disorder 
strengths. Focusing first on the inversion-symmetric case $\mathcal{D}_{\theta} = 0$ we find
\begin{equation}\label{RG_eqs}
\begin{aligned}
     \frac{\partial\mathcal{D}_{\phi}}{\partial l} =& \left(3 - 2K\right)\mathcal{D}_{\phi} , \phantom{..} 
        \frac{\partial{g_{\phi}}}{\partial l} = \left(2 - K\right)g_{\phi} \\ 
        \frac{\partial K}{\partial l} =& -\frac{4\pi^{2} K^{2}}{v_{F}^{2}}\left[ g^{2}_{\phi} + \mathcal{D}_{\phi}\right].
\end{aligned}
\end{equation} 
Here we can see that both $g_{\phi}$ and the disorder $\mathcal{D}_{\phi}$ flow to strong coupling for $K < 3/2$, and as $\mathcal{D}_{\phi} >0$ both renormalize the Luttinger parameter to values $K\left(l\right) < 1$. 
 In Sec. \ref{disorderDCI}, we discuss in detail disorder effects within the DCI phase for larger interactions and $K>2$.
\begin{figure*}
	\centering 
    \includegraphics[width =0.975\textwidth]{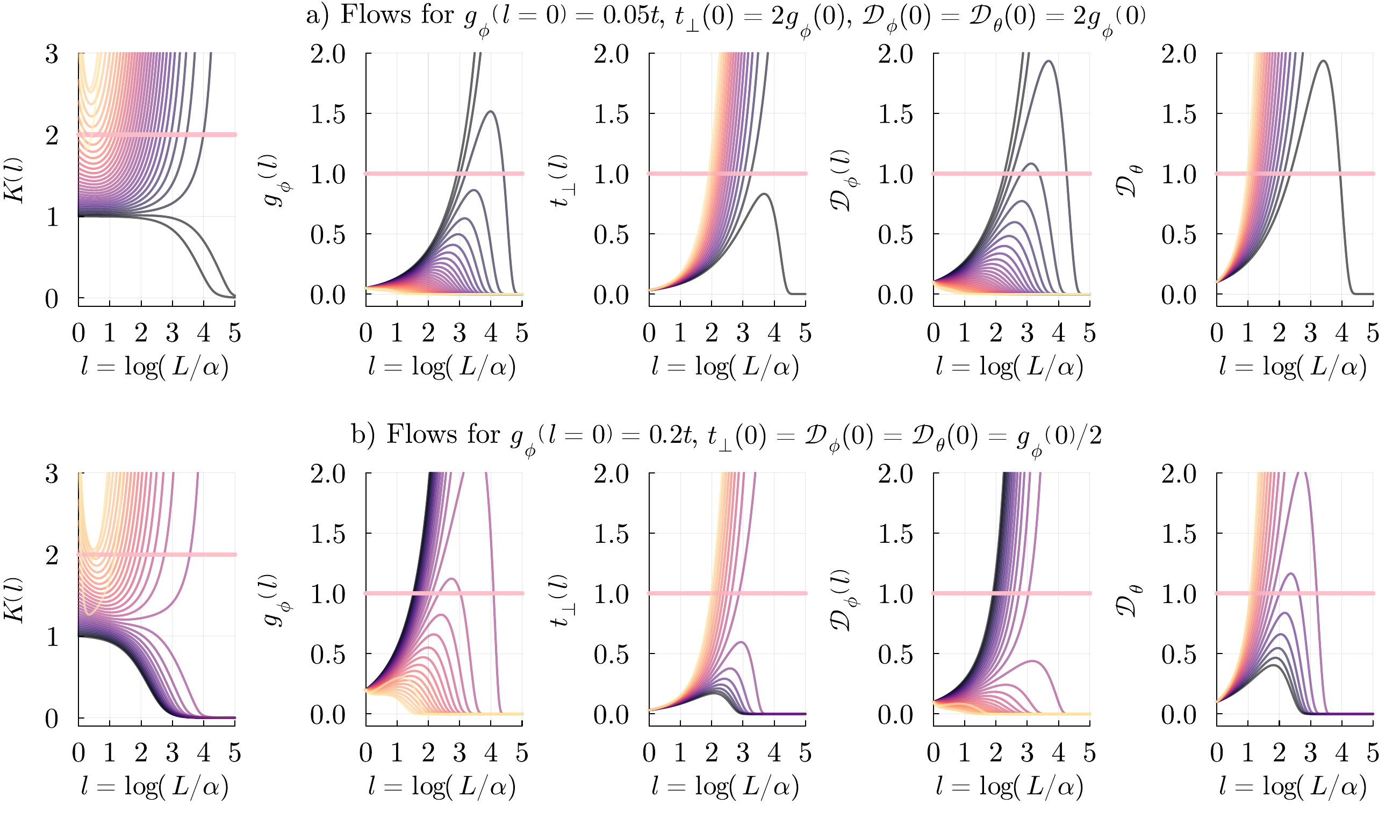}
	\caption{{\color{black}
	Numerically solved flow equations for $t_{\bot}\left(l = 0\right) = 0.1t$ and moderate disorder $\mathcal{D}_{\phi}(0) = \mathcal{D}_{\theta}(0) = 0.1t$ along the critical line for $g_{\phi}(0) = 0.05t$ and $g_{\phi}(0) = 0.2t$ respectively. Individual flows for $g/t \in \left[0.0, 3.0\right]$ with step-size $\delta g/t = 0.2$, and a color gradient \textit{dark} $\rightharpoonup$ \textit{light} indicating the progression of the flows with $g$. As in Fig. \ref{fig:flow_symmetric} above, the $x-$axis denotes the exponential length-scale $l$ with $L = \alpha e^{l}$, and the pink bands denote the $K = 2$ line and strong coupling values of the flows. } }
	\label{fig:flow2p}
\end{figure*}
The effects of inversion symmetry breaking disorder fluctuations in the boson picture are similar to those of an} \fremove{contribution $\mathcal{D}_{\theta} > 0$ are similar to those of an} inter-wire hopping term $t_{\bot}c^{\dagger,1}_{j}c^{2}_{j} + hc.$ which \fremove{This gave rise to an operator $\sin\left(2\theta\right)$ \footnote{The difference to \cite{del2023fractional} is that for $\mathcal{D}_{\theta} = 0$ a change of basis to $c^{\pm} \sim c^{1} \pm c^{2}$ resulted in a $\sim\cos\left(2\theta\right)$ term.}, which introduced a localization length $L^{*}$ for the critical modes in the DCI phase \cite{del2023fractional}.} \fremove{ The localization of the critical DCI modes} gave rise to a topological phase with shared edge modes between both wires.
The most general case for two proximized and interacting Kitaev wires includes the effects of an inter-wire hopping amplitude $t_{\bot}$. As we derive in the Appendix \ref{AppendixB1.2}, we then find the additional flow equations 
\begin{equation}\label{full_flow}
\begin{aligned}
       \frac{\partial K}{\partial l} =&
       -\frac{4\pi^{2}}{v^{2}_{F}}\left( K^{2}g^{2}_{\phi} - \tilde{t}_{\bot}^{2} + \mathcal{D}_{\phi}K^{2} - \mathcal{D}_{\theta}\right) \\ 
       \frac{\partial \tilde{t}_{\bot}}{\partial l} =& \left(2 - K^{-1}\right)\tilde{t}_{\bot} ,\phantom{.}  \frac{\partial \mathcal{D}_{\theta}}{\partial l} = \left(3 - 2K^{-1}\right)\mathcal{D}_{\theta},
\end{aligned}
\end{equation}
where $\tilde{t}_{\bot} = \frac{t_{\bot}}{\pi}$. We refer to Appendix \ref{AppendixB2} for more details on the derivation. \\
{\color{blue}\subsubsection{Disorder localization in the DCI phase\label{disorderDCI}}}
\lhreplace{ The DCI phase in the absence of disorder exists in the presence of strong interactions \cite{herviou2016phase} with a Luttinger parameter $ K > 2$. 
The interaction term $g_{\phi}$ then flows to zero, such that the fixed point is the Gaussian model in Eq. (\ref{Boson_Ham}) equivalent in structure to a free theory of the Majorana fermions in Eq. (\ref{Chiral_modes_link_one_wire}). 
In the presence of disorder, for $K>2$ then the symmetric disorder $\mathcal{D}_{\phi}$ also flows to zero.
Finally, if the wires are symmetric, then the bare value of the assymetric is simply $\mathcal{D}_{\theta}=0$.
The DCI phase is a gapless phase, which can be seen as a Majorana liquid.
Since the low-energy modes are delocalized on both wires (see Eq. (\ref{Chiral_modes_link_one_wire})), it is possible to have a topological phase that is in fact stable towards disorder in contrast to the QCP of a single wire. 
We have justified in Sec. \ref{Sec2} that in that case, the QCP is destabilized with any amount of disorder and the system is in a superposition
of topological and trivial phases when performing an ensemble average.}
\lhreplace{In fact}, the RG Eqs \eqref{RG_eqs} are  similar to those of the Mott/disorder transition \cite{giamarchi2004quantum, senechal2006theoretical, GiamarchiSchulz_loc1}, where the gap $g_{\phi}$ competes with $\mathcal{D}_{\phi}$ to provide the interaction strength --- \lhreplace{given by} whichever operator flows to \lhreplace{the} strong coupling first. 
\lhreplace{Here, } $g_{\phi}$ quantifies the distance to the critical line, and not \lhreplace{just} the repulsive interactions $g/t$. A diverging $g_{\phi}$ then only signals a departure from the DCI regime, ie. the \lhreplace{opening of the }$4$MF and trivial phases.

\freplace{We now present numerical solutions of the flow equations showcasing the above discussed behaviour of the interacting wires along the critical line $g_{\phi} = 0$. 
Due to the divergence of the bare parameter $K$ at large $g/t$, we are restricted to values $g/t < \pi$. To go beyond this threshold, we employ a DMRG analysis in short and periodic chains later on in Sec. \ref{Sec5}. Solutions of the RG flows are illustrated in Fig. \ref{fig:flow_symmetric} along the critical line $g_{\phi}(l = 0) = 0.01$ for symmetric order alone $\mathcal{D}_{\phi}(0) = 0.05t $ and $0.1t$ respectively. 
\lhremove{The plots show a progression of both $K\left(l\right)$ and $\mathcal{D}_{\phi}\left(l\right)$ as a function of the RG scaling $l = \log\left(L/\alpha\right)$ for interaction strengths $g/t \in \left[0.0,4.0\right]$ and step-size $\delta g/t = 0.2$, with a color gradient \textit{dark} to \textit{light} indicating the flows as a function of $g$.} 
The upper panels (a) in Fig. \ref{fig:flow_symmetric} show the effective suppression of inversion-symmetric disorder at large interactions $\mathcal{D}_{\phi}\left(l \longrightarrow \infty\right) = 0.0$. Conversely, if disorder becomes too strong, we find in panel (b) of Fig. \ref{fig:flow_symmetric}$b$ that for all interactions considered $K$ is renormalized below a value of $K < 3/2$. Hence, the operators associated to both $g_{\phi}$ and $\mathcal{D}_{\phi}$ are relevant and open a gap. In the presented case, the disorder fluctuations $\mathcal{D}_{\phi}$ flows to strong coupling before $g_{\phi}$ does, signaling a localization of the chiral DCI modes due to strong disorder.

Physically, we attribute the survival of the DCI phase to the pinning of charges due to strong interactions.
This is similar to the suppression of the charge disorder in the Mott phase of the one-dimensional Fermi-Hubbard chain, where 
at weak disorder and half-filling the charges are fixed by the interactions, and the spin degrees of freedom remain gapless.
}
\freplace{To go beyond this qualitative picture, a non-perturbative treatment of all the perturbations along the lines of \cite{senechal2006theoretical, Giamarchi_nonpert, Giamarchi_nonpert2} is necessary. We present a DMRG analysis of the DCI phase in short and periodic chains in Sec. \ref{Sec5}, which supports our qualitative picture derived from the RG Eqs (\ref{RG_eqs}) above.}\newpage

{\color{blue} \subsubsection{Stability of the $2$MF phase in the Kitaev ladder}}
\freplace{As we show in the Appendix \ref{AppendixB1.2}, the inter-wire hopping amplitude $t_{\bot}$ and symmetry breaking disorder fluctuations of the chemical potential $\mathcal{D}_{\theta}$ give rise to $\sin\left(2\theta\right)$ and $\cos\left(2\theta\right)$ contributions in the boson representation. }An analytical expression for the localization lengths for both operators can then be found from the RG equations in \eqref{RG_eqs} by solving with the ansatz $\partial_{l}K = 0$ \cite{giamarchi2004quantum}, ie. around a fixed point. When $\mathcal{D}_{\phi/\theta} \geq 0$, the only fixed point of these RG equations is characterized by $g^{*}_{\phi} = \gamma^{*}_{\theta} = \mathcal{D}_{\theta/\phi} =0$, and $K^{*} = A^{*}$ constant. Neglecting for a moment the symmetric disorder, we find close to $g_{\phi} = 0$ together with the ansatz $\tilde{t}_{\bot}\left(l\right) = \tilde{t}_{\bot,0}e^{\left(2 - K^{-1}\right)l}$ the length-scales associated to when $\tilde{t}_{\bot}\left(l\right) = 1$ and so on:
\begin{equation}\label{flow_analytical}
        L^{*}_{\tilde{t}_{\bot}} = \alpha\left(\frac{1}{\tilde{t}_{\bot,0}}\right)^{\frac{1}{2 - K^{-1}}}, \phantom{-} L^{*}_{\mathcal{D}_{\theta}} = \alpha\left(\frac{1}{\mathcal{D}_{\theta}}\right)^{\frac{1}{3 - 2K^{-1}}}
\end{equation}

Since $\tilde{t}_{\bot} = t_{\bot}/\pi$ enters the RG flow quadratically, for $t_{\bot}$ to flow to strong coupling before $\mathcal{D}_{\theta}$ we require $\mathcal{D}_{\theta} \ll t_{\bot}$, corresponding to a $2$MF topological phase with weak disorder. Then $\tilde{t}_{\bot}\sin\left(2\theta\right)$ \footnote{The difference to \cite{del2023fractional} is that for $\mathcal{D}_{\theta} = 0$ a change of basis to $c^{\pm} \sim c^{1} \pm c^{2}$ resulted in a $\sim\cos\left(2\theta\right)$ term.} pins the angle $\theta$ to $\pm \left(\pi/4 + n\pi\right)$, where $n$ is an integer. 
However, as symmetry breaking disorder contributes as a $\cos\left(2\theta\right)$ operator, the stability of the $2$MF phase is ensured since $\cos\left(\frac{\pi}{2} + 2n\pi\right) = \cos\left(\frac{\pi}{2}\right)$ which is then zero, when $\tilde{t}_{\bot}$ flows to strong coupling first. 
We assumed that $K$ remains constant - which is not guaranteed once we deviate from the fixed point. However, solving the flow equations shows a good agreement between the numerical and predicted $L^{*}$. 

\freplace{In Fig. \ref{fig:flow2p} above we showcase similarly as in the symmetry preserving case the effects of $t_{\perp}$ and $\mathcal{D}_{\theta}$, and their interplay with the symmetric terms $g_{\phi}$ and $\mathcal{D}_{\phi}$. As can be seen in the upper panels of Fig. \ref{fig:flow2p}$a$, the DCI phase is gapped first by the disorder fluctuations $\mathcal{D}_{\phi}$ and $\mathcal{D}_{\theta}$ at low interactions (dark lines). This corresponds to a disorder localized (trivial) phase. This is supported by a LL parameter $K\left(l \longrightarrow \infty\right) = 0$. However, at larger interactions (bright lines), the interwire hopping amplitude supresses the symmetric terms $g_{\phi}$ and $\mathcal{D}_{\phi}$, seen by a diverging LL paraeter $K$. Then, it is the competition between $t_{\bot}$ and $\mathcal{D}_{\theta}$ which determines if the wires are disorder localized, or in fact $2$MF topological when $t_{\bot}$ flows to strong coupling first \cite{del2023fractional}.}
Therefore, our RG analysis using the chiral fields defined in \cite{del2023fractional} further allows us to describe the stability of the $2$MF phase against local disorder. \\

{\color{blue} \section{DMRG analysis of disorder along the critical line}\label{Sec5}}


\freplace{Finally, we present a numerical study of the ground state properties of disordered and strongly interacting Kitaev wires. For this we use the DMRG algorithm \cite{PhysRevB.48.10345, PhysRevLett.69.2863, Schollw_ck_2005, Schollw_ck_2011} in the ITensors implementation in Julia \cite{Fishman_2022} to obtain the GS of the system up to a precision of $\leq 10^{-4}$ in the energy variance. Our analysis focuses both on local and global (topological) properties. The local properties are encoded by individual and local correlation functions as analysed in Sec. \ref{Sec2}, while the topological ones are determined from the $C$ marker and related observables. }

\begin{figure}[h!]
	\centering
	\includegraphics[width = \linewidth]{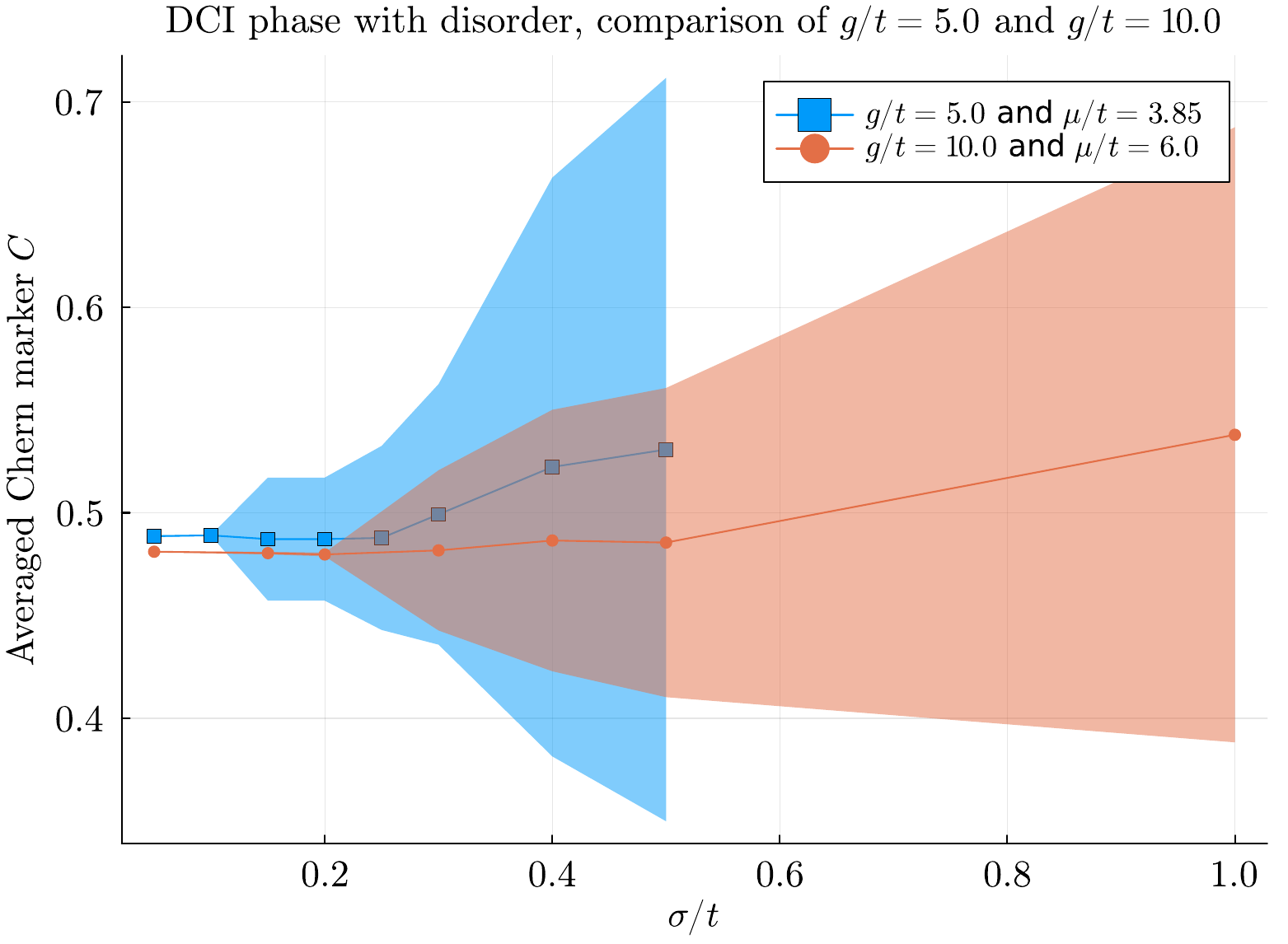}
	\caption{Comparison of the topological properties of the DCI phase in the presence of symmetric Gaussian disorder. DMRG results obtained for $N = 60$ chains with PBCs and $t = \Delta = 1.0$, and a GS energy variance of $\leq 10^{-4}$. Chern markers $C$ of wire "$1$" averaged over $M = 100$ samples of disorder. }
	\label{fig:DCI_g5vsg10}
\end{figure}

\freplace{In Fig. \ref{fig:DCI_g5vsg10} we see the averaged Chern markers within the expected (clean) DCI phase at $\mu/t = 3.85$ and $\mu/t = 6.0$ for $g/t = 5.0$ and $g/t = 10.0$ respectively shows a clear $\langle C\rangle  = 1/2$ plateau below critical disorder strengths $\sigma_{c}\left(g\right)$. The $\pm 1\sigma$ deviations are shown as a ribbon around the averaged values, and clearly show that disorder fluctuations are supressed below $\sigma_{c}$. This critical value is interaction dependent, which we can infer from the comparison of both figures at $g/t = 5.0$ and $g/t = 10.0$. This suggests a stabilization of the DCI phase against disorder at larger interactions, as predicted by the RG analysis in Sec. \ref{Sec4} before. \\
\begin{figure}[h!]
	\centering
	\includegraphics[width = \linewidth]{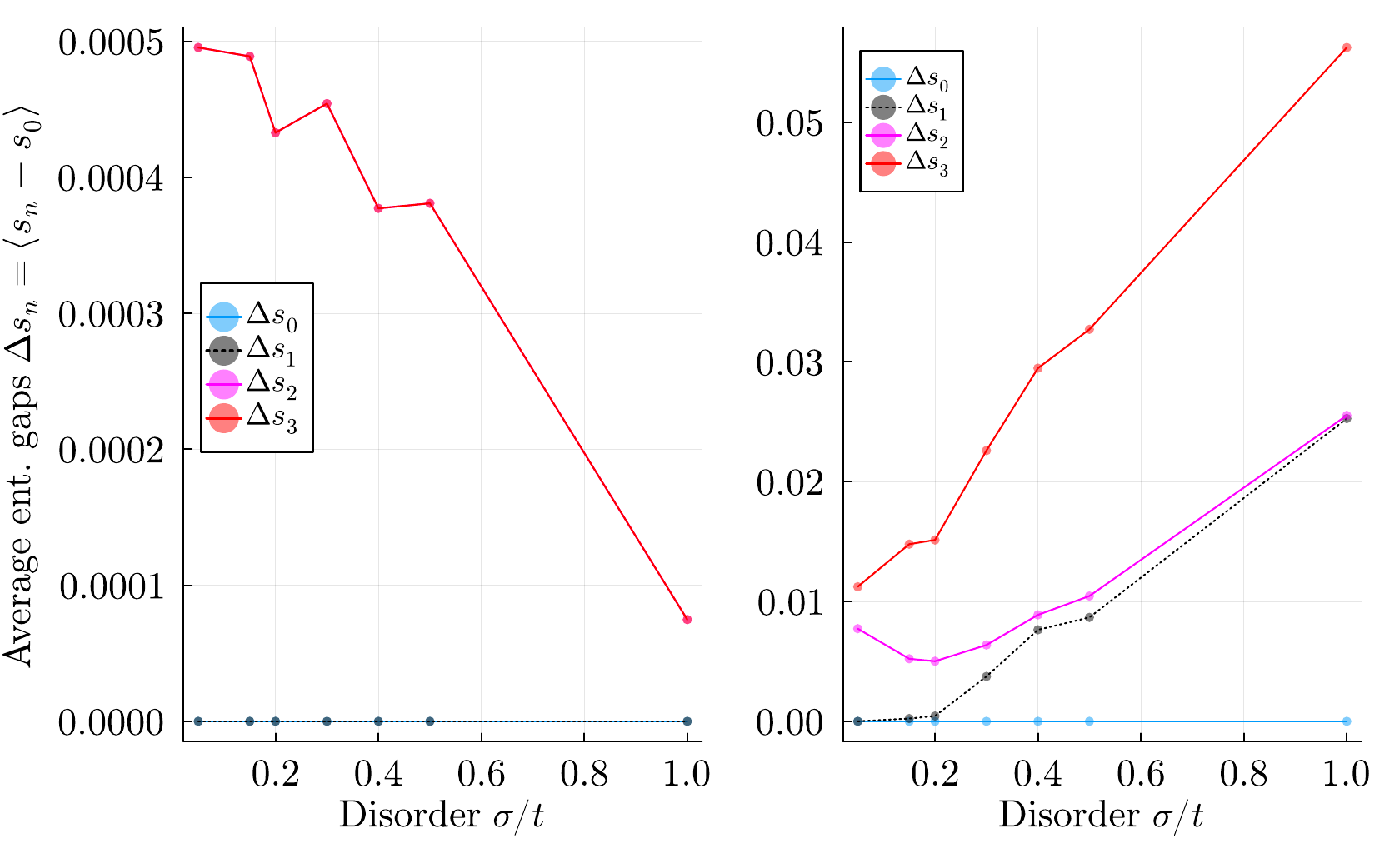}
	\caption{Four lowest lying eigenvalues of the entanglement spectra at different values of disorder strengths for $g/t = 10.0$ and at $\mu/t = 6.0$. DMRG performed for $t = \Delta = 1.0$ and $N = 60$ sites per wire with PBCs, and data points are averaged over $M = 100$ samples. }
	\label{fig:DCI_g10_ents}
\end{figure}

To verify that we are indeed within the DCI phase, we could extract disorder averaged central charges. However, the disorder fluctuations result in a near critical behavior, which is of the same order as the effects of using short and finite chains. Therefore, we instead compare the $\langle C\rangle  = 1/2$ plateau to the double-degeneracy of the entanglement spectrum \cite{EntanglementPOV} in Fig. \ref{fig:DCI_g10_ents} below. Therefore we conclude that our DMRG results in finite chains indeed captures the interaction induced stabilization against specific Gaussian disorder.}
{\color{black}As a last probe into the physical properties of the DCI phase, we extract the correlation functions in Fig. \ref{fig:correls} in various phases for the two wires model. We also show the behavior of the nearest-neighbor correlator in  Fig. \ref{fig:champagne_plots_PBC}.
The results for PBC nicely agree with the theoretical results for one wire at the QCP in Sec. \ref{Sec2}.
As we see in Fig. \ref{fig:champagne_plots_PBC}, there is a difference for the nearest-neighbor correlator 
 between PBCs and OBCs also at $N = 100$ sites per wire. \freplace{However, we remark that this difference is also of the same order in magnitude
as the standard deviation reported on the evaluation of disordered averaged correlators in Fig. \ref{fig:champagne_plots_PBC_smooth1}. Such a deviation may also stem from finite size effects. Furthermore, we observe a deviation between PBCs and the $1/3\pi$ QCP value, for $g/t = 5.0$ and $\mu/t = 3.8$. Since the correlation functions are not quantized we expect a deviation from the non-interacting value away from $g_{\phi} = 0$.}
Instead, we find a smooth behaviour of $\langle c^{\dagger}_{i+1}c_{i}\rangle$, as can be seen in Fig.  \ref{fig:champagne_plots_PBC_smooth1}. 
The averaged correlation functions across a single sample vary approximately smoothly across the disordered DCI phase as a function of $\mu/t$, such that the correlation functions could be used to indirectly quantify the effective deviation from $g_{\phi} = 0$ in the physical wire basis. Stability against disorder is suggested by the small standard deviation (error bars).

\begin{figure}[h!]
	\centering
	\includegraphics[width = 1.0\linewidth]{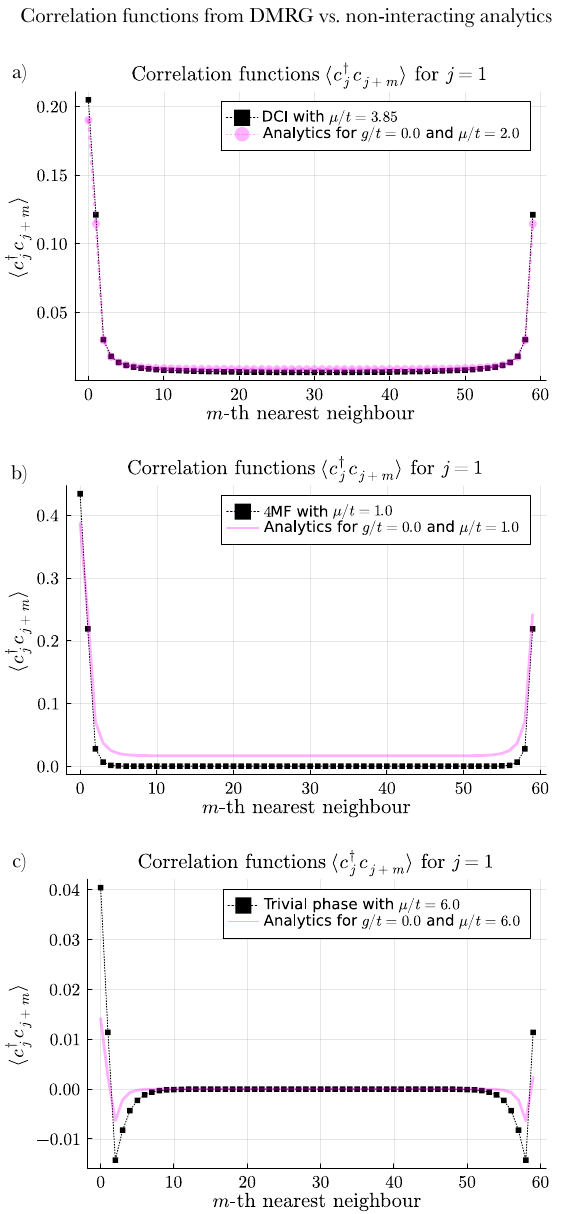}
\caption{{\color{black}Correlation functions obtained from DMRG for $N = 60$ sites per wire with PBCs, at $t=\Delta  =1.0$ and different values of the chemical potential. Pink lines correspond to the non-interacting analytical results of Sec. \ref{Sec2}. DMRG converged to precision of order of $\leq 10^{-4}$ in the energy variance, and cut-off value $\chi = 548$ for the maximum link-dimension.} }
	\label{fig:correls}
\end{figure}
\begin{figure}[h!]
	\centering
	\includegraphics[width = 1.0\linewidth]{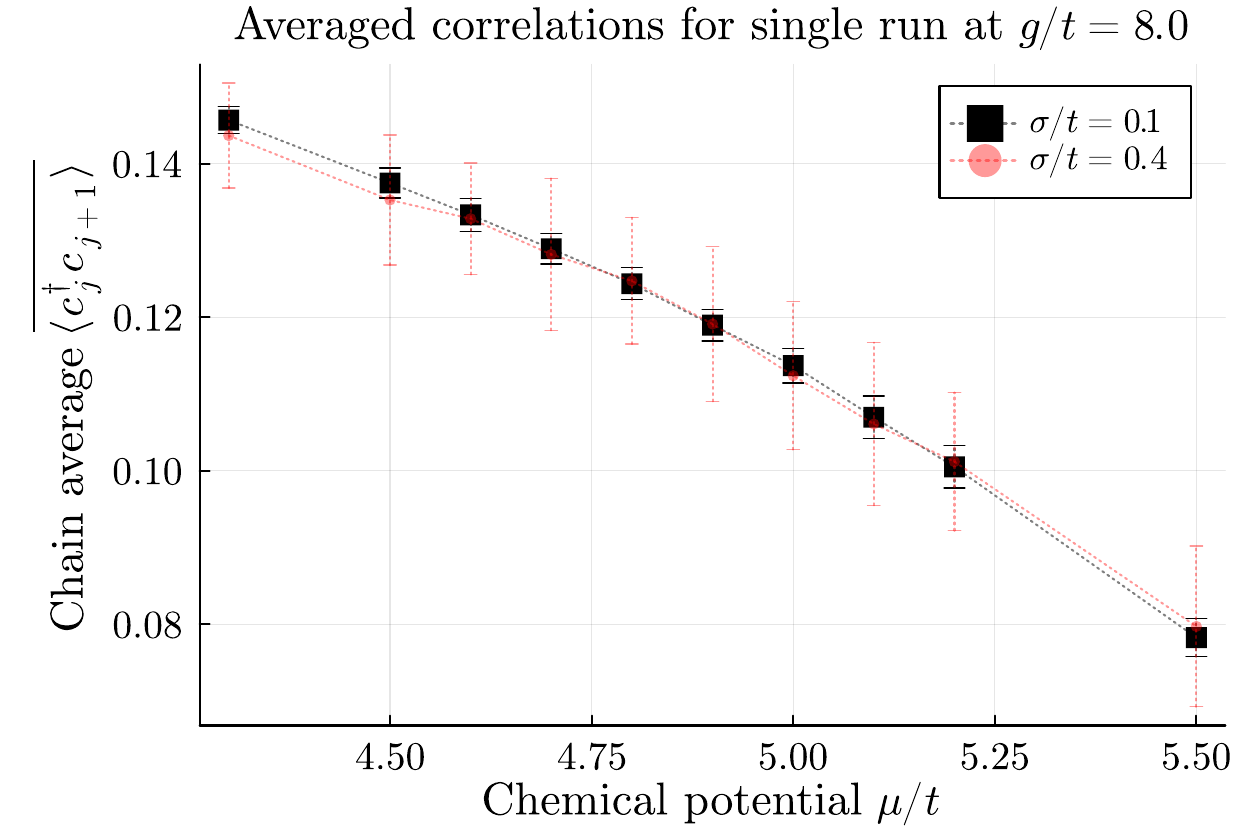}
\caption{Averaged correlation functions across wire $1$ for a single, representative realisation of disorder, for $g/t = 8.0$ and $t= \Delta = 1.0$. The correlation function $\langle c^{\dagger}_{j}c_{j+1}\rangle$ vary smoothly across the disordered DCI region. }
	\label{fig:champagne_plots_PBC_smooth1}
\end{figure}

In fact, as we see in Fig. \ref{fig:CorrelationFunction2Panels}, the correlation functions fluctuate around the average value with different degrees of spread depending on $\sigma/t$ and $g/t$. 
At the critical points for a single wire (ie. $g/t = 0.0$) the correlation functions were found to fluctuate more strongly as in the topological phase. 
Conversely, for $g/t = 5.0$ in panel (a) of Fig. \ref{fig:champagne_plots}, the correlation functions have less pronounced fluctuations for $\sigma/t = 0.4$.
Finally, at $g/t = 10.0$ in panel (b) we see for comparable disorder strengths that fluctuations remain constrained around the clean value $\approx 0.12$.
The exact spread depends on the specific disorder realisations. 
In summary, the behaviour of the correlations  captures the competition between repulsive Coulomb interactions $g/t$ and symmetry-preserving disorder $\mathcal{D}_{\phi}$, and provides a first simple numerical signature of stability against disorder at large interactions.}

{\color{blue}\section{Conclusions}\label{Sec5}}

{We have revisited the phase diagram of two interacting Kitaev wires in the presence of (local) Gaussian disorder in the form of a fluctuating chemical potential. We investigated the stability of the real-space topological marker $C$ previously established in \cite{del2023fractional}, and further developed the quantum field theoretical description of the \emph{doubly critical Ising} phase (DCI) far-from half-filling \cite{del2023fractional, herviou2016phase}, to include Gaussian disorder.

We find that the Chern marker follows the behaviour of other topological probes (topological gap, entanglement spectrum and edge-mode amplitudes) also when strong repulsive interactions are included between the chains. We verify a solid correspondence up to values of $\sigma/t \approx 0.5$ for $g/t = 2.0$. Beyond this, fluctuations in the correlation functions become significant, and the $C$ markers become increasingly blurry as a function of $\sigma/t$ and $\mu/t$. This establishes the Chern marker as a powerful tool to probe topological phases of interacting Kitaev wires, and offers a complementary approach to \cite{Gergs2016method}.

Our quantum field theoretic approach gives a comprehensive description near the critical region and far-from half-filling, where we demonstrated that interactions stabilize the DCI phase against inversion symmetric disorder. We further emphasize the usefulness of nearest-neighbor correlation functions to quantify the amount of localization physics at criticality
and to quantify topological properties both within the topological phase and also at the QCP (when $t=\Delta$). The QFT approach is not limited to the critical region, and we show how the $2$MF for non-zero inter-wire hopping $t_{\bot}$ \cite{del2023fractional} as well as the Mott transition close to half-filling remains stable for comparatively large values of $\sigma/t$. 

To the best of our knowledge, this is the first time an interaction protected transition from disordered to critical phase is found for topological superconducting wires. All in all, the reinforcing nature of repulsive Coulomb interactions on the topological region is a promising sign for potential applications in quantum computing and information storage. 
\lhreplace{
The transport properties of this exotic gapless regime remain a topic for further studies, as well as the impact of light-matter coupling in the presence of a superconducting cavity.
We also observed large boundary effects in the DCI regime: the existence and nature of edge modes --- if any-- would be another interesting perspective to explore.
%
Finally, similar correlators in fermionic models have been related to the quantum Fisher information, which was recently measured \cite{Brookhaven}.
}\\

{\color{blue}\section*{Acknowledgements}}
Frederick del Pozo would like to thank the Ecole Doctorale of IP Paris for continued financial support throughout the project. In addition, Frederick del Pozo would like to thank in particular Sariah al Saati and Ephraim Bernhardt for insightful discussions and comments throughout the project. Karyn Le Hur thanks the Deutsche Forschungsgemeinschaft (DFG),  German Re-search Foundation under Project No.  277974659, for financial support. Olesia Dmytruk acknowledges support from the ERC Starting Grant, Grant Agreement No.~101116525.\\

\appendix

{\color{blue}\section{Kitaev wire and correlation functions}\label{AppendixA}}

\freplace{In the Appendix we derive and expand on the various two-point correlation functions introduced in Sec. \ref{Sec2}. This includes first a detailed derivation and comparison of the dominant nearest-neighbor correlation functions also in the presence of disorder, as well as a subsequent derivation of longer-range correlations near the quantum critical points. }\\

\subsubsection{Dominant nearest neighbor correlation functions as an indicator for disorder localization}
We introduced the Hamiltonian of the interacting Kitaev wire in Eq.~\eqref{eq:KitaevH}. 
Here, we present a quick derivation of the correlation functions
$\langle c^{\dagger}_{j+1} c_j\rangle$ and $\langle c_j^\dag c_{j+1}^\dag\rangle$ for the BCS ground state. 
We include weak interaction effects and reproduce the linear evolution of the QCP with the interaction $V$ \cite{Schuricht} through a simple mean-field theory which modifies linearly the hopping and pairing terms while maintaining the form of the BCS wavefunction. 
We emphasize that this approach is purely qualitative.
In the Luttinger liquid formalism, it corresponds to the inclusion of an interaction of the form $\frac{V}{\pi^2}(\partial_x\phi_{\sigma})^2$ \cite{giamarchi2004quantum} followed by discarding the renormalization of the Luttinger parameter, so that only the Fermi velocity (or identically $t$) is renormalized linearly with $V$.
Introducing the hopping channels $\langle c_j^\dag c_{j+1}\rangle,\ \langle c_{j+1}^\dag c_{j}\rangle$ and pairing channels $\langle c_j^\dag c_{j+1}^\dag\rangle, \ \langle c_{j+1}c_{j}\rangle$,
we write the mean-field Hamiltonian
\begin{align}
	&H_{K}^{MF} = -\sum_{j=1}^{N}\mu c_j^\dag c_j - \sum_{j=1}^{N-1}\left(c_j^\dag c_{j+1} (t+ V\langle c_{j+1}^\dag c_{j}\rangle) + \text{h.c.}\right) \notag\\
	&+\sum_{j=1}^{N-1}\left(c^\dag_j c^\dag_{j+1} (\Delta + V \langle c_{j+1}c_j\rangle) + \text{h.c.}\right).
	\label{eq:KitaevHInteractions2}
\end{align}
We naturally introduce the notations
\begin{align}
    t_\mathrm{eff} = t+ V\langle c_{j+1}^\dag c_{j}\rangle, \ \
    \Delta_\mathrm{eff} = \Delta + V \langle c_{j+1}c_j\rangle.
\end{align}
The BCS ground state takes the general form
\begin{equation}
\label{BCSgroundstate}
|\psi_{GS}\rangle = {\cal R} \prod_{0<k<\frac{\pi}{a}}(u_k+v_k c^{\dagger}_k c^{\dagger}_{-k})|0\rangle,
\end{equation}
where ${\cal R} = (\delta_{\mu<-2t_\mathrm{eff}}+(1-\delta_{\mu<-2t_\mathrm{eff}})c^{\dagger}_0)\cdot
(\delta_{\mu<2t_\mathrm{eff}}+(1-\delta_{\mu<2t_\mathrm{eff}})c^{\dagger}_{\pi})|0\rangle$ \lhreplace{and $\delta$ is the Kronecker symbol (with the lowerscript $\mu<-2t_\mathrm{eff}$ emphasising that this is $1$ if the constraint is satisfied and 0 otherwise).}
\lhremove{\color{black}  the symbol $\delta_{\mu<2t_\mathrm{eff}}$
means that this (the $\delta$ Kronecker function) is equal to one when $\mu<2t_\mathrm{eff}$ and zero otherwise and the symbol 
$\delta_{\mu<-2t_\mathrm{eff}}$ is equal to one when $\mu<-2t_\mathrm{eff}$ and zero otherwise.} We can absorb the phase of $\Delta$ through the gauge transform $c_j \rightarrow e^{i\frac{\varphi}{2}}c_j$ and therefore consider $\Delta$ and $\Delta_\mathrm{eff}$ real without loss of generality in the following.
Finally, if we define
\begin{align}
\label{mathequations}
\cos \theta_k &= \frac{- \mu - 2 t_\mathrm{eff}\cos k}{E_k}, \phantom{...}
\sin \theta_k = \frac{2 \Delta_\mathrm{eff}\sin k}{E_k}\\
E_k &= \sqrt{ (\mu + 2 t_\mathrm{eff}\cos k)^2  + 4 \Delta_\mathrm{eff}^2 \sin^2 k },
\end{align}
we have $u_k = \cos \frac{\theta_k}{2}$ and $ 
v_k = i \sin \frac{\theta_k}{2}$.
Here, we naturally assume that the lattice spacing $a$ is set to unity.
Standard computations lead to
\begin{align}
\label{correlationfunction}
\langle \psi_{GS}| c^{\dagger}_j c_{j+1} |\psi_{GS}\rangle &= \frac{1}{\pi} \int_{0}^{\pi} dk 
|v_k|^2 \cos k \\
&= - \frac{1}{2 \pi} \int_{0}^{\pi} dk 
\cos \theta_k \cos k
\end{align}
and similarly $
\langle \psi_{GS}| c^{\dagger}_j c^{\dagger}_{j+1} |\psi_{GS}\rangle = \frac{1}{2\pi}\int_{0}^{\pi}  \sin \theta_k \sin k dk$.
We focus first on the limit where $\mu = 0$ and $\Delta_{\mathrm{eff}} = t_{\mathrm{eff}}$.
Straightforward replacement gives $\theta_k = \pi - k$ and 
\begin{equation}
    \langle \psi_{GS} | c_j^\dag c_{j+1} 	|\psi_{GS}\rangle = \langle \psi_{GS} | c_j^\dag c_{j+1}^\dag 	|\psi_{GS}\rangle = \dfrac{1}{4}.
\end{equation}
The line $\mu = 0$ and $\Delta_{\mathrm{eff}} = t_{\mathrm{eff}}$ is therefore stable under self-consistent mean-field.
{\begin{figure}[h!] 
	\includegraphics[width=0.75\linewidth]{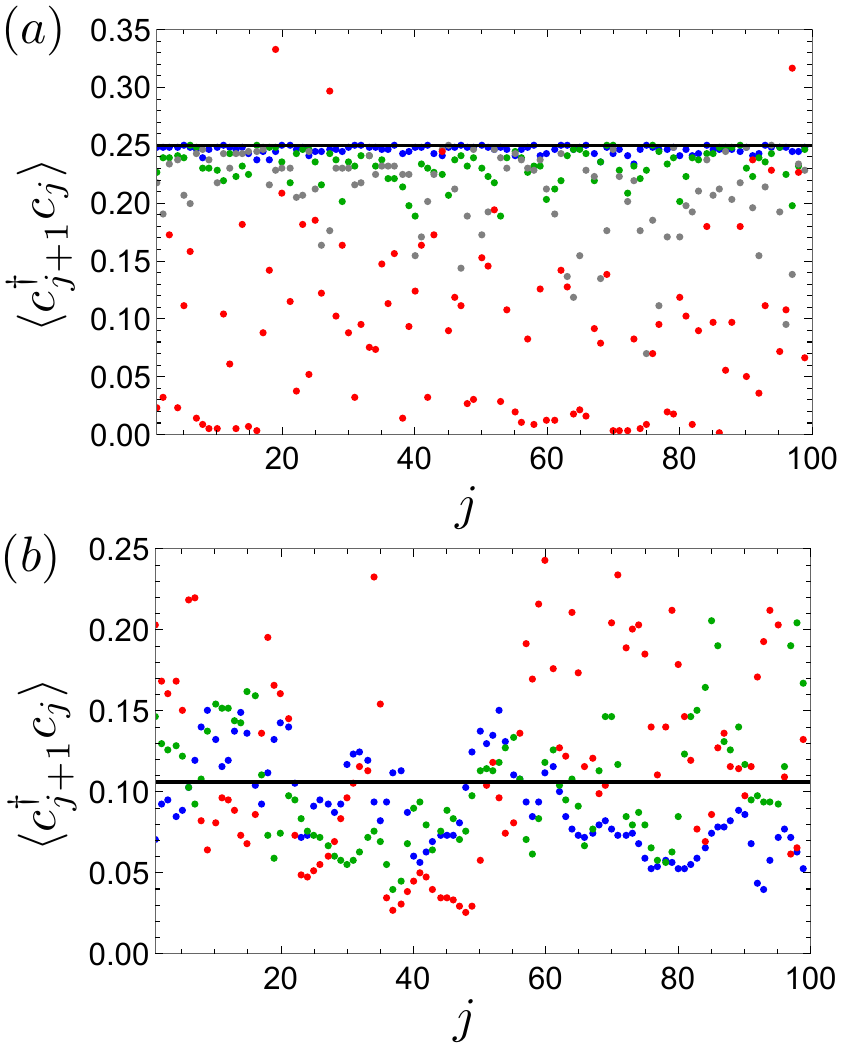}
	\caption{Correlation function $\langle c_{j+1}^\dag c_j\rangle$ as a function of the lattice site $j$ for the disordered Kitaev wire with PBCs. (a) For the clean wire in the topological phase with $\mu = 0$  we find that $\langle c_{j+1}^\dag c_j\rangle = 1/4$ (black solid line). Blue dots correspond to a single realization of the Gaussian disorder with $\sigma = 0.5 t$. Green dots correspond to $\sigma = t$, gray dots correspond to $\sigma = 1.5t$, and red dots correspond to $\sigma = 5t$. (b) For the clean wire at the QCP with $\mu = -2t$ we obtained that $\langle c_{j+1}^\dag c_j\rangle = 1/(3\pi)$ (black solid line). Blue dots correspond to a single realization of the Gaussian disorder with $\sigma = 0.5 t$, green dots correspond to disorder with $\sigma = 0.7t$, and red dots correspond to disorder with $\sigma = t$. We fixed $\Delta = t = 1$, and $N = 100$.  }
	\label{fig:CorrelationFunction2Panels}
\end{figure}

If we start on this line, the interactions simply renormalize $t$ and $\Delta$ in $t + \frac{V}{4}$ and $\Delta$ in $\Delta + \frac{V}{4}$, so that only the velocity is renormalized.
The system can still be diagonalized in terms of Majorana fermions \cite{hur2022topological,herviou2016phase}. In this sense, fixing $t=\Delta$ in the calculation above remains legitimate in the presence of (weak) interactions. Following these facts, in the present article, for the clean case we will apply the form of Eq. (\ref{correlationfunction}) which is gauge independent for the correlation function and is equal to $\frac{1}{4}$ if $t=\Delta$.
The quantum critical points are reached when $\mu=\pm 2t_\mathrm{eff}$, where we can similarly evaluate the correlation function.
Let us consider $\mu = 2 t_\mathrm{eff}$ and $\Delta_{\mathrm{eff}} = t_{\mathrm{eff}}$ for simplicity. Then $\theta_k$ is equal to $\frac{\pi}{2} - \frac{k}{2}$, and 
\begin{equation}
    \langle \psi_{GS} | c_j^\dag c_{j+1} 	|\psi_{GS}\rangle = \frac{1}{3\pi}\sim 0.106,
\end{equation}
whilst $ \langle \psi_{GS} | c_j^\dag c_{j+1}^\dag 	|\psi_{GS}\rangle = \dfrac{2}{3\pi}$.
This computation gives us a good intuition on the shift of the effective critical chemical potential under the addition of the interaction: $\mu_\mathrm{eff} = 2(t + \frac{1}{3 \pi} V)$ for $\Delta = t - \frac{V}{3 \pi}$.
\begin{figure}[h!] 
	\includegraphics[width=0.8\linewidth]{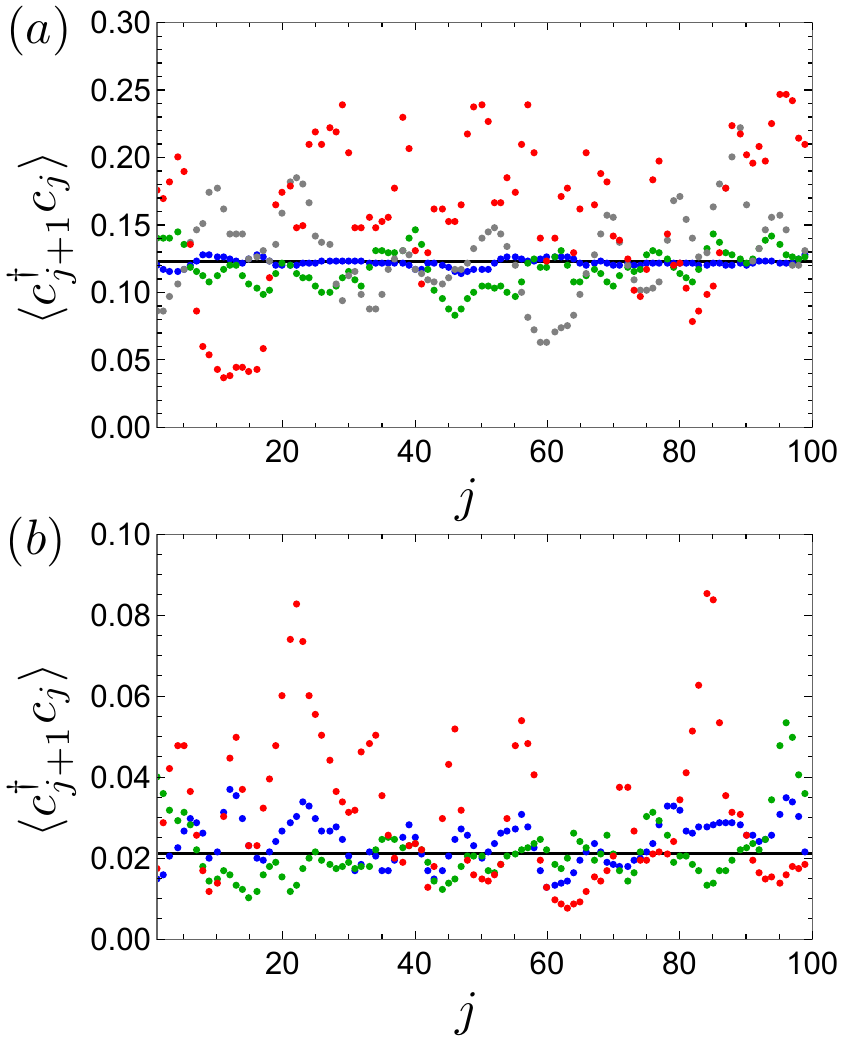}
	\caption{ (a) Correlation function $\langle c_{j+1}^\dag c_j\rangle$ as a function of the lattice site $j$ for the disordered Kitaev wire close to the phase transition ($\mu = -1.95t$) with PBCs. For the clean wire $\langle c_{j+1}^\dag c_j\rangle \approx 0.123$ (black solid line). Single realization of the Gaussian disorder with $\sigma = 0.05 t$ (blue), $\sigma = 0.2t$ (green), $\sigma = 0.5t$ (gray) $\sigma = t$ (red). 
 (b) Correlation function $\langle c_{j+1}^\dag c_j\rangle$ as a function of the lattice site $j$ for the disordered Kitaev wire in the trivial phase ($\mu = -3t$) with PBCs. For the clean wire we obtained that $\langle c_{j+1}^\dag c_j\rangle \approx 0.021$ (black solid line). Single realization of the Gaussian disorder with $\sigma = 0.5 t$ (blue), $\sigma = 0.7t$ (green), and $\sigma = t$ (red).  
 Other parameters are chosen as $\Delta = t = 1$, and $N = 100$.
 }
	\label{fig:FigCorrelationCloseTransition}
\end{figure}
The evaluation of the mean-field critical line for $t = \Delta$ can readily be done numerically, without significant qualitative change.
We can also evaluate these correlation functions analytically as a function of the chemical potential within the same approach which agrees with the black solid line in Figs. \ref{fig:FigCorrelationCloseTransition}. These figures also show the evolution of the correlation functions as a function of disorder strength in the vicinity of the QCP with ED. We can verify the numerical results of Fig. \ref{fig:CorrelationFunctionDeltaNeqT} with the analytical formula
\begin{equation}
\label{formulatDelta}
       \langle \psi_{GS}|c^{\dagger}_j c_{j+1} |\psi_{GS}\rangle =
    \frac{1}{2\pi}\int_0^{\pi}
    \frac{e^{-ik} sgn(t\cos k +\frac{\mu}{2}) }{\sqrt{1+\frac{\Delta^2 \sin^2 k}{\left(t\cos k +\frac{\mu}{2}\right)^2}}}dk.
\end{equation}
\begin{figure}[h!] 
	\includegraphics[width=0.8\linewidth]{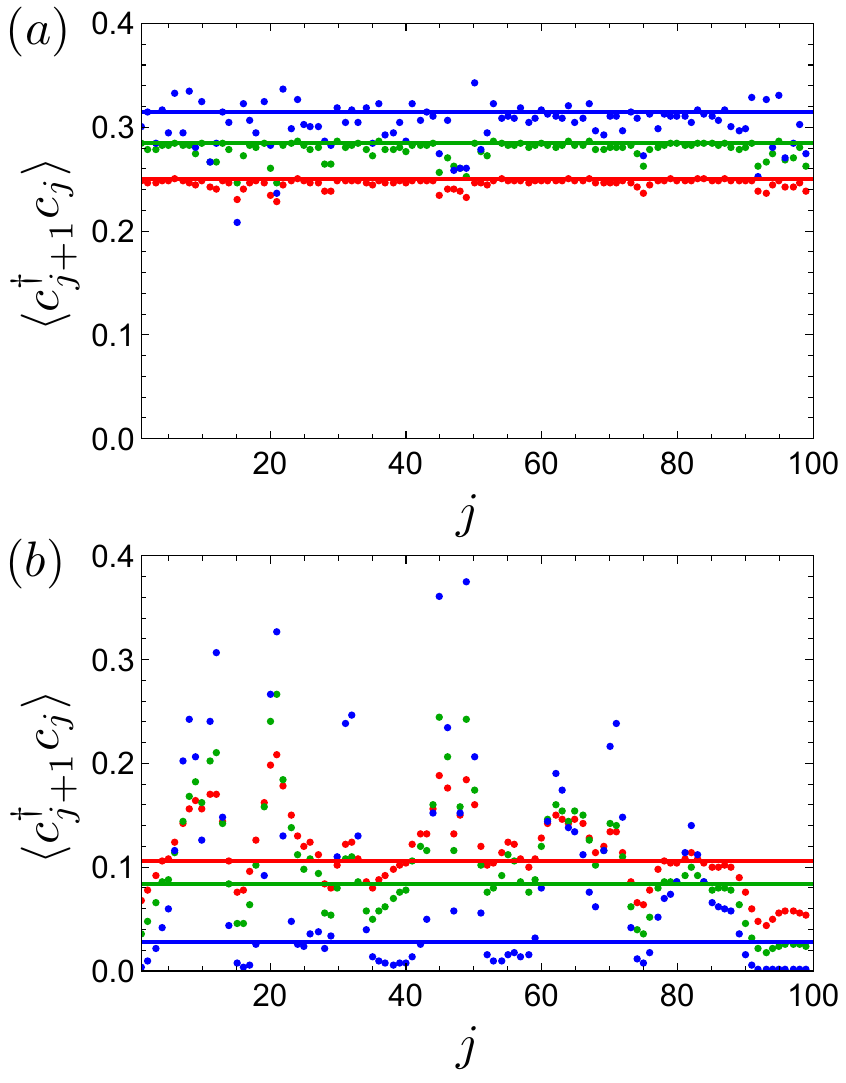}
	\caption{
 Correlation function $\langle c_{j+1}^\dag c_j\rangle$ as a function of the lattice site $j$ for the disordered Kitaev wire with PBCs calculated for a single realization of the Gaussian disorder with $\sigma = 0.5t$ for $\Delta = t$ (red), $\Delta = 0.5t$ (green) and $\Delta = 0.1t$ (blue). (a) In the topological phase with $\mu = 0$ for the clean wire case $\langle c_{j+1}^\dag c_j\rangle = 1/4$ (horizontal red solid line) for $\Delta/t = 1$. The value of the correlation function increases with decreasing $\Delta/t$: $\langle c_{j+1}^\dag c_j\rangle \approx 0.285$ (horizontal green solid line) for $\Delta/t = 0.5$ and $\langle c_{j+1}^\dag c_j\rangle \approx  0.315$ (horizontal blue solid line) for $\Delta/t = 0.1$.
 (b) At the QCP  $\mu = -2t$ for the clean wire case $\langle c_{j+1}^\dag c_j\rangle = 1/(3\pi)$ (horizontal red solid line) for $\Delta/t = 1$. The value of the correlation function decreases with decreasing $\Delta/t$: $\langle c_{j+1}^\dag c_j\rangle \approx 0.084$ (horizontal green solid line) for $\Delta/t = 0.5$ and $\langle c_{j+1}^\dag c_j\rangle \approx  0.028$ (horizontal blue solid line) for $\Delta/t = 0.1$, and $N = 100$ sites.
}
	\label{fig:CorrelationFunctionDeltaNeqT}
\end{figure}

\subsubsection{Non-local correlation functions in the Chern number}
\label{nonlocalChernnumber}

{\color{black} As discussed in Sec. \ref{Sec2}, the continuous version of the Chern number $C$ in Eq. (\ref{eq:Cnumber}) is given by the infinite sum
\begin{equation}
\label{eq:Cthermo}
C = \sum_{i=-\infty}^{+\infty} \langle c^{\dagger}_j c_{j+ 2i + 1} + 
c_{j + 2i + 1}^{\dagger} c_j \rangle,
\end{equation}
accounting for the thermodynamic limit. For $\mu=0$ and $t=\Delta$ the two terms coming from $c_{j+N-1}=c_{j-1}$ with PBCs in Eq. (\ref{eq:Cnumber}) are also equivalently encoded with $c_{j+2i+1}$ where $i=-1$ for OBCs. 
We also verify Eq. (\ref{eq:Cthermo}) for $\mu=0$ and $\frac{\Delta}{t}\rightarrow 0$. }

\freplace{It is noteworthy that at the QCPs we find $C = \pm 1/2$. This can equally be} interpreted as a correspondence with a half sphere (or half Skyrmion). 
The correlator $\langle c^{\dagger}_{j+1} c_j\rangle$ becomes equal to $\frac{1}{3\pi}$ simply from the fact that $C=\frac{1}{2}$ is equivalent to a half-sphere in the analogy between a spin-1/2 particle and the $2\times 2$ matrix description of the BCS model 
\cite{hur2022topological,del2023fractional}. Precisely, for the specific situation of $t=\Delta$ and with the formalism introduced in Appendix~\ref{AppendixA}, we obtain a simple relation between the polar angle $\theta$ on the sphere and the wave-vector $k$ such that $\cos\theta_k=\sin\frac{k}{2}$ and 
$\sin\theta_k=\cos\frac{k}{2}$ which is equivalent to $\theta_k=-\frac{k}{2}+\frac{\pi}{2}$. For simplicity, the lattice spacing is fixed to unity. We can then establish a simple form
for the local correlation function on the sphere. The correlation function reads
\begin{equation}
    \langle \psi_{GS} | c^{\dagger}_j c_{j+1} | \psi_{GS}\rangle=\frac{1}{\pi}\int_0^{\frac{\pi}{2}} d\theta_k\sin^2\theta_k \cos\theta_k.
\end{equation}
Including a shift $\theta_k' = \theta_k - \frac{\pi}{2}$ we obtain the identity $\langle \psi_{GS} | c^{\dagger}_j c_{j+1} | \psi_{GS}\rangle=
\frac{1}{3\pi}\left(\cos(0)-\cos \left(\frac{\pi}{2}\right)\right)$. \freplace{Similarly, we find that at $\mu=0$ the angle $\theta_{k} = k$ and hence 
the correlation functions simply become 
\begin{equation}
\begin{aligned}
       \langle \psi_{GS} |c^{\dagger}_{j}c_{j+1}|\psi_{GS}\rangle =& -\frac{1}{2 \pi}\int_{0}^{\pi}\text{d}\theta_{k}\cos^{2}\left(\theta_{k}\right) \\ =& \frac{1}{4\pi}\left[ x - \frac{1}{2}\sin\left(2x\right)\right]_{x = 0}^{x = \pi} \\ 
       =& \frac{1}{4\pi}\pi  = \frac{C}{4}.
\end{aligned}
\end{equation}
The latter identification with the Chern number $C$ stems from the fact that $\langle c^{\dagger}_{j+m}c_{j}\rangle  =0$ for $m \neq -1,1$, and hence we have the strict identification from \eqref{eq:Cthermo} that 
\begin{equation}
    C  = \sum_{\tau = \pm 1}\langle c^{\dagger}_{j}c_{j + \tau} + \text{hc.}\rangle  = 4 \langle c^{\dagger}_{j}c_{j+1}\rangle.
\end{equation}

Thus, in the specific and special limits $\mu = 0$ or $\mu/t = \pm 2t$ for $t = \Delta$, we find an intriguing correspondence between the nearest-neighbor correlation functions and topological invariants of the non-interacting Kitaev wire.}

{\color{blue}\section{RG approach for the disordered wires}}\label{AppendixB1}

{\color{blue}\subsection{One and two wires at weak interactions}\label{disorderflow}}

Here, for completeness, we prove the stability of the topological p-wave superconducting phase in the presence of Gaussian disorder in the bulk, through RG flows equations, for one and two wires. 
We use the notations of Giamarchi and Schulz related to our previous analysis of Ref. \cite{herviou2016phase} for the two (clean) wires at weak interactions $g$. 
This approach implicitly assume that the Fermi velocity $v_F=2ta\sin k_F$ is of the order of $t$
such that the kinetic term will be described through the Luttinger liquid theory. The
chemical potential is then fixed such that $-2t\ll \mu\ll 2t$.

Suppose we begin with two wires when $g=0$. We can introduce the fermion operators $c^{\dagger}_{\pm 1,\sigma=1,2}(x)$ similarly as in Ref. \cite{herviou2016phase}. The symbol 1,-1 refers to right- and left-moving particles and $\sigma=1,2$ labels the wire. Here, we apply the continuum description such that $c_j\rightarrow \frac{c(x)}{\sqrt{a}}$ with $a$ the lattice spacing. 
The fluctuating potential modelling disorder effects
can be included through a term $V_{\sigma}(x)c^{\dagger}_{1,\sigma}(x)c_{-1,\sigma}(x)+h.c.$. Within the bosonization technique, this gives rise to an additional term in the Lagrangian $\frac{V_{\sigma}}{2\pi \alpha}\cos(2\phi_{\sigma}(x))$, with the short-distance cutoff $\alpha$. This can be chosen proportional (or equal to) the lattice spacing $a$, but we keep this second scale explicit for completeness.  In the presence of a Gaussian probability distribution with $V\sim \delta\mu$
then from Gaussian integrals tricks, this gives rise to an effective term in the action \cite{giamarchi2004quantum}
\begin{equation}
-\int dx\int d\tau\int d\tau' \frac{{\mathcal D}_{\sigma}}{(2\pi \alpha)^2}\cos[2\phi_{\sigma}(x,\tau)-2\phi_{\sigma}(x,\tau')].
\end{equation}
We suppose here that the forward scattering term is independent and does not modify the RG flow
dominated by the backward scattering term $\mathcal{D}$. We also have the identity $\overline{V_{\sigma}(x)V_{\sigma}(x')}=\mathcal{D}_{\sigma}\delta(x-x')$, when averaging over disorder realizations, which implies that the action has only one integral in space and two integrals in time. The parameter $\mathcal{D}_{\sigma}$ plays a similar role as the variance of the Gaussian distribution that was also defined as $\sigma$ for the lattice calculation.
Since for weak disorder, various distributions can be approximated through a Gaussian distribution the results below will be generally applicable in the presence of weak disorder.

To derive the RG flows for the disorder amplitude $\mathcal{D}_{\sigma}$ and the pairing term $\Delta_{\sigma}$, we present here a simple approach through the invariance of the partition function. If we include the pairing terms and disorder perturbatively, we can apply the Luttinger liquid results \cite{giamarchi2004quantum}
\begin{eqnarray}\label{GreensFunc}
\langle [\phi_{\sigma}(r)-\phi_{\sigma}(0)]^2\rangle_0 &=& K_{\sigma}F(r) \nonumber \\
\langle [\theta_{\sigma}(r)-\theta_{\sigma}(0)]^2\rangle_0 &=& K_{\sigma}^{-1}F(r),
\end{eqnarray}
with $r=(x,\tau)$ and
\begin{equation}\label{Ffunc}
F(r) = \frac{1}{2}\log \frac{x^2 + (|\tau|^2 +\alpha)^2}{\alpha^2}
\end{equation}    
which dominates the behavior of correlation functions within the quadratic Luttinger liquid theory
denoted through the symbol $_0$. For simplicity, the velocity is set to unity such that space and
imaginary time have the same dimension. To have the action dimensionless, this requires that
$\mathcal{D}_{\sigma}$ has the same dimension as the inverse of a length. It is then useful to introduce the dimensionless variable $\tilde{\mathcal{D}}_{\sigma}=\mathcal{D}_{\sigma}\alpha$. 
The Luttinger parameter $K_{\sigma}$ is equal to one for free electrons. In general, Coulomb interactions effects will modify the value of $K_{\sigma}$ as well as the velocity. 

The invariance of the partition function $\mathcal{Z}(\alpha)=\mathcal{Z}(\alpha')$ or of the physics when modifying slightly the short-distance cutoff can be revealed through the identification of correlation functions for the Luttinger liquid theory
\begin{equation}
\langle e^{i2(\phi_{\sigma}(x,\tau)-\phi_{\sigma}(x,\tau')}\rangle_0=e^{-2\langle (\phi_{\sigma}(x,\tau)-\phi_{\sigma}(x,\tau')^2\rangle_0}.
\end{equation}
This is equivalent to say that the physics at long wavelength would not depend on the definition of the short-distance cut-off (comparable to the lattice spacing) being $\alpha$ or $2\alpha$. $\mathcal{Z}(\alpha)=\mathcal{Z}(\alpha')$ then leads to the relation $
\frac{\tilde{\mathcal{D}}_{\sigma}(\alpha')}{\tilde{\mathcal{D}}_{\sigma}(\alpha)} = 
\left(\frac{\alpha'}{\alpha}\right)^{3-2K_{\sigma}}$, such that for
a small variation $\alpha'=\alpha e^{dl} = \alpha(1+dl)$ we find \cite{GiamarchiSchulz_loc1,giamarchi2004quantum}
\begin{equation}
\frac{\partial\tilde{\mathcal{D}}_{\sigma}}{\partial l} = (3-2K_{\sigma})\tilde{\mathcal{D}}_{\sigma}.
\end{equation}
We can then introduce the pairing term $\Delta$ in each wire which gives rise to a ter $
\frac{2\Delta\sin(k_F)}{\pi \alpha}  \cos(2\theta_{\sigma})$
in the Lagrangian. We can then introduce a dimensionless pairing function $\tilde{\Delta}=\Delta \alpha$
such that reproducing the same calculation as above leads to $
\frac{\partial\tilde{\Delta}^2}{\partial l} = \left(4-\frac{2}{K_{\sigma}}\right)\tilde{\Delta}^2$.
The factor $4$ instead of $3$ comes from the fact that when developing the action to second order in
$\Delta$ then we have two integrals on time and two integrals on space. This is also equivalent to
\begin{equation}
\frac{\partial\tilde{\Delta}}{\partial l} = \left(2-\frac{1}{K_{\sigma}}\right)\tilde{\Delta}.  
\end{equation}

For free electrons since $K_{\sigma}=1$ (or weak interactions within each wire compared to the pairing term) then we observe that disorder and pairing term will flow similarly to strong couplings such that we can justify the stability of the p-wave superconductor until $\tilde{\mathcal D}_{\sigma}(\alpha)<\tilde{\Delta}_{\sigma}(\alpha)$ which is similar to say $\sigma<\Delta$ within the lattice definitions. The equation for $\tilde{\Delta}$ can be solved analytically introducing the length $L$ of the wire such that $\alpha'/\alpha=\log(L/\alpha)$. The parameter
$\tilde{\Delta}$, which is supposed to be small at short length scales allowing us to develop
the action to second-order in perturbation theory, will then reach the strong-coupling fixed point
when $\tilde{\Delta}(L^*)\sim 1$ (which is defined to be of the order of the kinetic term where
we set the velocity being equal to one). Therefore, at the length scale 
$L^* \sim \alpha\left(\tilde{\Delta}(\alpha)\right)^{\frac{1}{2-\frac{1}{K_{\sigma}}}}$ this is the same as if $\Delta^*=\Delta(L^*)\sim t$. In the equation for $L^*$, $\Delta(\alpha)=\Delta$ refers to the bare value of the pairing term at short distance. 

Therefore, this implies that in the bulk, under the RG flow, 
the physics of the p-wave SC will remain very similar as long as $\sigma<\Delta$ even if we begin
with $\Delta<t$. At the low-energy fixed point or long-wavelength limit, this is then similar
to pin the phase $\theta_{\sigma}$ to a minimum of the cosine potential. In this sense, the
$\phi_{\sigma}(x,\tau)$ variable fluctuates strongly such that effectively $\langle \cos(2\phi_{\sigma}(x,\tau)\rangle\sim 0$ similarly as if we would average equally on all angles
for $\phi_{\sigma}\in [0;2\pi]$. We can then summarize that as long as $\sigma<\Delta$, then
the correlation function will remain (almost) identical $\langle c^{\dagger}_{j+1}c_j\rangle\sim \frac{1}{4}$. This conclusion remains also valid in the presence of an asymmetry in the pairing terms in the two wires as long as $(\Delta_1(\alpha),\Delta_2(\alpha))>\mathcal{D}$. However, when increasing the Coulomb interaction in a wire, resulting in decreasing $K_{\sigma}$, we observe that localization may develop in the weak-$\Delta$ limit.

Including a weak Coulomb interaction $g$ between wires, 
it is then judicious to introduce the modes
$\phi_{\pm}=\frac{1}{\sqrt{2}}(\phi_1\pm \phi_2)$ and similarly for the $\theta_{\pm}$ variables
mixing the two wires. We can then proceed along the lines of Ref. \cite{herviou2016phase} re-writing the term in the Lagrangian associated to the pairing term $\Delta$ as
\begin{equation}
\frac{\Delta^{(1)}}{\alpha^2}\cos(\sqrt{2}\theta_+)\cos(\sqrt{2}\theta_-).    
\end{equation}
The disorder term in each wire gives a term of the form
\onecolumngrid
\begin{equation}\label{dis_half_filling}
\frac{\tilde{\mathcal{D}}_{\sigma}}{(2\pi)^2\alpha^3}\cos\left(2\left(\frac{\phi_+(x,\tau')\pm \phi_-(x,\tau')}{\sqrt{2}} -\frac{\phi_+(x,\tau)\pm \phi_-(x,\tau)}{\sqrt{2}}\right)\right).
\end{equation}
\twocolumngrid
in the Lagrangian. This gives rise to the RG equations
\begin{eqnarray}
\frac{\partial\tilde{\mathcal{D}}_{\sigma}}{\partial l} &=& (3-K_+ - K_-)\tilde{\mathcal{D}}_{\sigma} 
\nonumber \\
\frac{\partial\tilde{\Delta}^{(1)}}{\partial l} &=& \left(2-\frac{1}{2}(K_+^{-1} +K_-^{-1})\right)\tilde{\Delta}^{(1)},
\end{eqnarray}
with the Luttinger parameters \cite{herviou2016phase}$
K^{-1}_{\pm} = \sqrt{1\pm \frac{g}{\pi v_F}}$.
For small $g$ values, these RG equations can be simplified
\begin{equation}
\frac{\partial \tilde{\mathcal{D}}_{\sigma}}{\partial l} = \left(1-\frac{3}{4}\frac{g^2}{v_F^2}\right)
\tilde{\mathcal{D}}_{\sigma} \phantom{..} \text{and} \phantom{..}
\frac{\partial \tilde{\Delta}^{(1)}}{\partial l} = \left(1+\frac{1}{8}\frac{g^2}{v_F^2}\right)
\tilde{\Delta}^{(1)}.
\end{equation}
To summarize, for weak interactions between wires, we observe that this now helps the pairing term to grow to strong couplings faster compared to disorder effects. \\

{\color{blue}
\subsection{Mott transition close to half-filling}\label{AppendixB1.2}}

The bosonic representation of two interacting Kitaev wires close to half-filling has been derivedƒ prior in \cite{herviou2016phase}. Here we summarise the main results of this derivation. 

Bosonization is performed close to the Fermi-momentum $k_{F}$, which at half-filling results in a Fermi velocity $v_{F} = 2t\sin\left(k_{F}\right) \approx 2t$. The individual bosonic modes on each wire can be combined into the bonding/anti-bonding basis $\pm$, and results in the following Hamiltonian 
\begin{equation}
\begin{array}{r}
H=\sum_{\varepsilon= \pm} \int d x \frac{v_{F, \varepsilon}}{2 \pi}\left(K_{\varepsilon}\left(\partial_x \theta_{\varepsilon}\right)^2+K_{\varepsilon}^{-1}\left(\partial_x \phi_{\varepsilon}\right)^2\right) \\
+\frac{g_{\varepsilon}}{\alpha^2} \cos \left(2 \sqrt{2} \phi_{\varepsilon}\right)-\frac{\Delta_{\varepsilon}^{(2)}}{\alpha^2} \cos \left(2 \sqrt{2} \theta_{\varepsilon}\right) \\
+\frac{\Delta^{(1)}}{\alpha^2} \cos \left(\sqrt{2} \theta_{+}\right) \cos \left(\sqrt{2} \theta_{-}\right) .
\end{array}
\end{equation}
The bare values of $g_{\pm}$ are $\mp g/(2\pi^2)$, whilst $\Delta^{1}(0) = \frac{4\Delta \alpha}{\pi}$. The bare values of the $\pm$ Fermi velocities and Luttinger parameters are given by 
\begin{equation}
\begin{aligned}
v_{F, \pm} = v_F \sqrt{1 \pm \frac{g}{\pi v_F}}, \phantom{...}
K_{ \pm}(0)=\frac{1}{\sqrt{1 \pm \frac{g}{\pi v_F}}}
\end{aligned}
\end{equation}
The superconducting pairing is reflected in the pairing $\Delta^{(1)}_{\pm}$, whilst the mixed $\pm$ pairing $\Delta^{(1)}$ is zero in the bare Hamiltonian with non-zero contributions generated through the RG flows 
\begin{equation}\label{Flows_half-filling}
\begin{aligned}
\frac{d K_{ \pm}}{d l} & =-\frac{2 \pi^2 g_{ \pm}^2}{v_{F, \pm}^2} K_{ \pm}^2+\frac{2 \pi^2\left(\Delta_{ \pm}^{(2)}\right)^2}{v_{F, \pm}^2}+\frac{\pi^2\left(\Delta^{(1)}\right)^2}{4 v_{F,+} v_{F,-}} \\
\frac{d g_{ \pm}}{d l} & =\left(2-2 K_{ \pm}\right) g_{ \pm} \\
\frac{d \Delta^{(1)}}{d l} & =\left(2-\frac{1}{2}\left(K_{+}^{-1}+K_{-}^{-1}\right)\right) \Delta^{(1)} \\
\frac{d \Delta_{ \pm}^{(2)}}{d l} & =\left(2-2 K_{ \pm}^{-1}\right) \Delta_{ \pm}^{(2)}+\frac{2 \pi^2\left(\Delta^{(1)}\right)^2}{v_{F, \mp}}.
\end{aligned}
\end{equation}
By summing the disorder contribution in \eqref{dis_half_filling} over the wire labels $\sigma$, and fixing $\mathcal{D}_{1} = \mathcal{D}_{2} \equiv \mathcal{D}$. The resulting flow equations are equivalent to those in \cite{GiamarchiSchulz_loc1}: In addition to a renormalization of the Luttinger parameters $\frac{d K_{\pm}}{d l}=-\frac{u_{\pm}}{2 u_{-}} K_{\pm}^2 \mathcal{D}(l)$ disorder also renormalizes the Fermi velocities due to the space/time assymetry \cite{GiamarchiSchulz_loc1}
\begin{equation}
\begin{aligned}
 \frac{d v_{F, \pm}}{d l}=-\frac{v^{2}_{F, \pm}}{2v_{F,-}}K_{\pm}\mathcal{D}
\end{aligned}
\end{equation}
As this effect is linear in the disorder strength, it can be neglected in the flows $\sim \mathcal{D}$ in \eqref{Flows_half-filling}, but not in the flows of $\Delta^{(1)}$ and $\Delta^{(2)}$. \\

The derivation of the above RG equations was technically done in \cite{GiamarchiSchulz_loc1} for a disorder term 
\begin{equation}
\begin{aligned}
\sim &\cos\left(\sqrt{2}\phi_{+}\left(r\right)\right)\cos\left(\sqrt{2}\phi_{+}\left(r'\right)\right)\\ &\times\cos\left(\sqrt{2}\left(\phi_{-}\left(r\right) - \phi_{-}\left(r'\right)\right)\right),
\end{aligned}
\end{equation} 
and not as in our case 
\begin{equation}
    \cos\left(\sqrt{2}\left(\phi_{+}\left(r\right) - \phi_{+}\left(r'\right)\right)\right)\cos\left(\sqrt{2}\left(\phi_{-}\left(r\right) - \phi_{-}\left(r'\right)\right)\right)
\end{equation}
However, 
the derivation of the flow equations in the Appendix of \cite{GiamarchiSchulz_loc1} can be applied to our disorder term, since the only non-zero terms in $\langle \prod_{i} e^{iA_{i}\phi\left(r_{i}\right)}\rangle$ are those which satisfy the neutrality condition $\sum_{i}A_{i} = 0$ \cite{giamarchi2004quantum}. This means, that the resulting terms in the RG calculation of both operators coincide, and we obtain equivalent RG equations as in \cite{GiamarchiSchulz_loc1}.
\\

{\color{blue}
\subsection{Far-from half-filling and in the DCI phase}\label{AppendixB2}}
Here we discuss the effects of disorder on two superconducting wires along the critical line $g_{\phi} = 0$ \cite{herviou2016phase, del2023fractional}, around which at large enough interactions the DCI phase opens up. We include on-site disorder directly into the QFT description around the critical line, found to be a $c = 1$ chiral Luttinger liquid \cite{del2023fractional} with Sine-Gordon like interaction in the bosonized form \cite{herviou2016phase, del2023fractional}. We first derive the representation of the disorder Hamiltonian in the chiral basis, and subsequently bosonize after which we a disorder averaging along the lines of \cite{GiamarchiSchulz_loc1} gives the flow-equations quantifying the competition between disorder and interactions. Finally, we offer a discussion on the physical picture. Refermionizing the final action gives insight into the way that disorder and interactions compete and give rise to the final phase diagram shown schematically in Fig. \ref{fig:PhaseDiagramlowdis}. 
\\
\subsubsection{Disorder in the chiral fermion basis}

We consider an on-site disorder potential of the form $\mu^{\sigma}\left(x\right) = \mu + \delta\mu^{\sigma}\left(x\right)$ with $\delta\mu^{\sigma}\left(x\right)$ given by a Gaussian distribution centered around $\mu = 0$ and with variance $\sigma^{2}$ on each wire. We may write this equivalently in terms of symmetric disorder $2\mathcal{V}= \delta\mu^{1} + \delta\mu^{2}$ across both wires, and asymmetric disorder $2\nu = \delta\mu^{1} - \delta\mu^{2}$. For $\mathcal{V}$, the fluctuations enter the tight-binding Hamiltonian as the following (in the continuum limit) \cite{del2023fractional} density term $n_{x} = \sum_{\sigma}c^{\dagger,\sigma}_{x}c^{\sigma}_{x}$
\begin{equation}\label{disorder_0}
    H_{dis} = \int_{x} \mathcal{V}\left(x\right)n_{x} \text{d} x = \int_{x} \mathcal{V}\left(x\right) \Gamma^{\dagger}_{x}\Theta_{x} \text{d} x  + \text{hc.}.
\end{equation}
Introducing the chiral modes $\psi_{R/L}$ defined in \eqref{Chiral_modes_link_one_wire} we find that disorder takes the well known Giamarchi-Schulz \cite{GiamarchiSchulz_loc1} cosine potential after bosonization \cite{del2023fractional}
\begin{equation}\label{disorder_bosonized}
H_{dis,\mathcal{V}} = \int_{x} \text{d}x \ \frac{\mathcal{V}\left(x\right)}{2\pi\alpha}\cos\left(2\phi\left(x,t\right)\right). 
\end{equation}

Similarly, the asymmetric disorder $\nu\left(x\right)$ enters as a term proportional to $\Delta n_{x} = n^{1}_{x} - n^{2}_{x}$, ie.
\begin{equation}
   H_{dis, \nu} = \int_{x} \nu\left(x\right) \Delta n_{x}\text{d}x = \int_{x}\Gamma^{\dagger}_{x}\Theta^{\dagger}_{x} \text{d}x+ \text{hc.} 
\end{equation}
Similarly after bosonization we find \cite{del2023fractional}
\begin{equation}\label{disorder_bosonized}
H_{dis,\nu} = \int_{x} \text{d}x \ \frac{\nu\left(x\right)}{2\pi\alpha}\cos\left(2\theta\left(x,t\right)\right). 
\end{equation}

\subsubsection{Disorder average for RG equations}

The disorder averaging can now be done along the same lines as in appendix \ref{AppendixB1}, ie. using the replica-field method \cite{GiamarchiSchulz_loc1, senechal2006theoretical}.  For a general and broad range of disorder realisations, the fluctuations $\mathcal{V}\left(x\right)$ and $\nu\left(x\right)$ are uncorrelated distributions, such that any cross terms $\langle \mathcal{V}\left(x\right)\nu\left(x\right)\rangle = 0$.
Defining the Gaussian covariance matrices $\mathcal{D}_{\phi/\theta}^{ab} = \delta^{ab}\mathcal{D}_{\phi/\theta}$ respectively, disorder averaging with the replica method produces the following two operators in the ``effective" action 
\begin{equation}
\begin{aligned}
       \delta S_{dis} &= \sum_{\epsilon = \pm}\int_{x, \tau, \tilde{\tau}} \Bigg( \frac{\mathcal{D}_{\phi}^{ab}}{\alpha^{2}} \cos\left(2\phi_{a}\left(x,\tau\right)+2\epsilon \phi_{b}\left(x,\tilde{\tau}\right)\right) \\
    +& \frac{\mathcal{D}_{\theta}^{ab}}{\alpha^{2}} \cos\left(2\theta_{a}\left(x, \tau\right)+ 2\epsilon \theta_{b}\left(x, \tilde{\tau}\right)\right)\Bigg)\text{d}x\text{d}\tau\text{d}\tilde{\tau}
\end{aligned}
\end{equation}
In the case of homogeneous (or symmetry-preserving) disorder $\nu = 0$, we therefore obtain the well known Giamarchi-Schulz type disorder potential \cite{GiamarchiSchulz_loc1}. In the presence of a de-tuning $\nu$, both potentials compete against eachother in a similar way as $t_{\bot}$ and $g$ in the Kitaev ladder case \cite{del2023fractional}. Due to the two fields $\phi$ and $\theta$ being conjugate variables to each other, the scaling dimensions of the $\cos$ operators are inverted. Therefore, we find for the flow equations of $g_{\phi}$ as before and $\mathcal{D}_{\phi}$ and $\mathcal{D}_{\theta}$ the following expressions
\begin{equation}
\begin{aligned}
    \frac{\partial\mathcal{D}_{\phi}}{\partial l} =& \left(3 - 2K\right)\mathcal{D}_{\phi} , \phantom{..} 
        \frac{\partial{\mathcal{D}_{\theta}}}{\partial l} = \left(3 - \frac{2}{K}\right)\mathcal{D}_{\theta},
\end{aligned}
\end{equation}
together with $\frac{\partial g_{\phi}}{\partial l} = \left(2 - K\right)g_{\phi}$.
Finally, we also find additional contributions to the flow of the Luttinger parameter $K$. In the disorder-free case these arise from contributions of $g_{\phi}$ \cite{herviou2016phase, del2023fractional, giamarchi2004quantum}, and can be derived from the correlation functions $C\left(r-\tilde{r}\right) = \langle e^{i\phi\left(r\right)} e^{-i\phi\left(\tilde{r}\right)}\rangle$, as corrections to the ``free" correlation function $F\left(r-\tilde{r}\right)$ introduced in \eqref{GreensFunc} and \eqref{Ffunc} above. As $\mathcal{D}_{\phi}$ already contributes an operator $\sim \cos\left(2\phi\right)^{2}$, such a contribution to $K$ is of linear order in $\mathcal{D}_{\phi}$. Therefore, following a tedious but straight-forward calculation, beautifully summarized in \cite{giamarchi2004quantum}, we find 
\begin{equation}
        \frac{\partial K}{\partial l} = -\frac{4\pi^{2} K^{2}}{v_{F}^{2}}\left[ g^{2}_{\phi} + \left(\mathcal{D}_{\phi} - \frac{\mathcal{D}_{\theta}}{K^{2}}\right)\right]
\end{equation}
The $\mathcal{D}_{\theta}$ term is accounted for in a very similar fashion, however it naturally dresses the inverse Luttinger parameter $K^{-1}$. However, using $\partial_{l}K^{-1} = -K^{-2}\partial_{l}K$ we readily find that $\mathcal{D}_{\theta}$ dresses $K$ linearly and pushes it to strong coupling. Thus, $\mathcal{D}_{\phi}$ and $g_{\phi}$ compete with $\mathcal{D}_{\theta}$ in the RG flow of $K$. However,  all of these considerations are made from a perturbative approach of disorder and are therefore suited as an indicative analysis on the interacting phase diagram of two Kitaev wires. For a more detailed dive into the various localized phases and scaling properties of the RG equations, we envision an analysis along the lines outlined in \cite{senechal2006theoretical} and the Gaussian Variational Method. This was beyond the scope of the current study and describes an interesting project to be addressed in the future. 
{\color{black}
\subsubsection{Inter-wire hopping}
Due to the occurrence of both $n^{1} \pm n^{2}$ terms in the presence of disorder, a change of basis into the bonding/anti-bonding basis \cite{del2023fractional} is no longer useful. We therefore bosonize the effects of an inter-wire hopping term 
\begin{equation}\label{tperp}
    H_{\perp} = t_{\bot} \sum_{j} c^{\dagger,1}_{j}c^{2}_{j} + \text{hc.},
\end{equation}
directly using the definitions in Appendix \ref{AppendixB2}. This leads to an additional operator $H_{\perp} = -\frac{t_{\bot}}{\pi \epsilon} \int_{x}\text{d}x \sin\left(2\theta\right)$.
This expression is slightly different from \cite{del2023fractional}, where a change of basis to the bonding/anti-bonding basis was made\footnote{This leads to a $\cos\left(2\theta\right)$ expression and $t_{\bot}/2$ pre-factor.} The flow of $\sin\left(2\theta\right)$ is the same as that of $\cos\left(2\theta\right)$ due to only a global phase difference, such that we find the final flow equations 
\begin{equation}\label{full_flow}
\begin{aligned}
       \frac{\partial K}{\partial l} =&
       -\frac{4\pi^{2}}{v^{2}_{F}}\left( K^{2}g^{2}_{\phi} - \tilde{t}_{\bot}^{2} + \mathcal{D}_{\phi}K^{2} - \mathcal{D}_{\theta}\right) \\ 
       \frac{\partial \tilde{t}_{\bot}}{\partial l} =& \left(2 - K^{-1}\right)\tilde{t}_{\bot} ,\phantom{.}  \frac{\partial \mathcal{D}_{\theta}}{\partial l} = \left(3 - 2K^{-1}\right)\mathcal{D}_{\theta}
\end{aligned}
\end{equation}}

{\color{blue}
\section{DMRG and other numerical specifics }} \label{AppendixC}

In the following appendix we provide further details on the numerical results presented for the two interacting and disordered wires. We first present a summary of the DMRG procedure used throughout the paper, and then provide further details on the topics of convergence and system sizes. We then present the additional figures and discussions regarding the topological marker in the presence of disorder. Finally, we also briefly address the numerical implementations of the flow equations and associated critical length scales. 
{\color{blue}
\subsection{DMRG algorithm and convergence}}
Throughout this paper we use the GS approximation scheme known as \emph{density matrix renormalization group} (DMRG) method, in the case of two coupled and interacting (disordered) wires. Reformulated in terms of Matrix Product States (MPS), it has found its way to become a widely respected tool, notably with implementations in C++ and Julia through the ITensors package \cite{ITensors}. Developed by White in the early nineties \cite{PhysRevLett.69.2863, PhysRevB.48.10345}, the procedure efficiently computes the lowest lying eigenstates of (quasi-)one dimensional systems. For a pedagogical introduction we refer the reader to \cite{Chiara2006DensityMR}, whilst the standard review with respect to the MPS formulation is \cite{Schollw_ck_2011}. 

Our current implementation for the two interacting Kitaev wires was developed in the context of \cite{del2023fractional}, and we summarize here the key features discussed therein.
A natural formulation of the DMRG method is achieved for one-dimensional chains, such that for (quasi-)two-dimensional systems like our coupled wires we recast the ladder onto a $1$d chain. This map is not unique, and specific choices are made strategically dependent on the model. With both inter-chain hopping $t_{\bot}$ and interaction $\sim g$, we choose a zig-zag pattern as in \cite{del2023fractional}. In the case of PBCs, an additional link was added between the last and first sites.  
Lastly, we note the importance of resolving the individual parity sectors explicitly, as the total parity of the individual wires is conserved in the absence of inter-wire hopping. Therefore the physical GS is in an explicit mixture $|GS\rangle = \alpha |GS,1\rangle + \beta |GS,2\rangle$ of both parity sectors $(P_{1},P_{2}) = (0,1)$ and $(1,0)$, and as was shown in \cite{del2023fractional} the DCI phase with PBCs is characterized by the many body entangled state, ie. $\alpha = \beta = \frac{1}{\sqrt{2}}$. 

{\color{black}For the convergence of the GS, we typically chose a GS energy variance cut-off value of $10^{-5}$ or $10^{-6}$, irrespective of the phase. However, gapped phases converge both quicker and ``better'', with higher degrees of convergence reached deep in the topological and trivial phases. We generally increased the maximum available link-dimension of the MPS gradually, until both convergence was achieved and the final link-dimension of the quantum state did not saturate the upper bound. We usually chose a threshold of $\chi_{GS}/\chi_{max} = 0.8$. To avoid getting stuck in a non-global minimum we also included an error term \cite{Fishman_2022}, which was exponentially reduced as the link-dimension was increased. A final run with error $\epsilon = 0$ was performed at the end. For more information on system-sizes and convergence of the entanglement spectrum, we refer the reader to the following sections, as well as our previous study in \cite{del2023fractional}, where a more detailed analysis is performed also with comparison to the Majorana fermion edge modes.}

{\color{blue}
\subsection{OBC Chern markers for two wires}}
Throughout the paper, we present results obtained from DMRG for finite wires of length $L = Na$ per chain. That means, we perform DMRG for a combined length of $2N$, setting $a = 1$ computationally. We encountered numerous differences between open boundaries and PBCs. As pointed out in \cite{del2023fractional}, the $C$ marker is derived from the momentum space representation, and thus we expect deviations from the perfectly quantized values already at $g/t = 0.0$ for finite chains. The second is, that with OBCs, the Kitaev wire presents two edge-modes per wire, localized at either extremities of the chain. These have a non-zero overlap, and thus lift the GS degeneracy due to a splitting of the energies. 

\begin{figure}[h!]
	\centering
	\includegraphics[width = \linewidth]{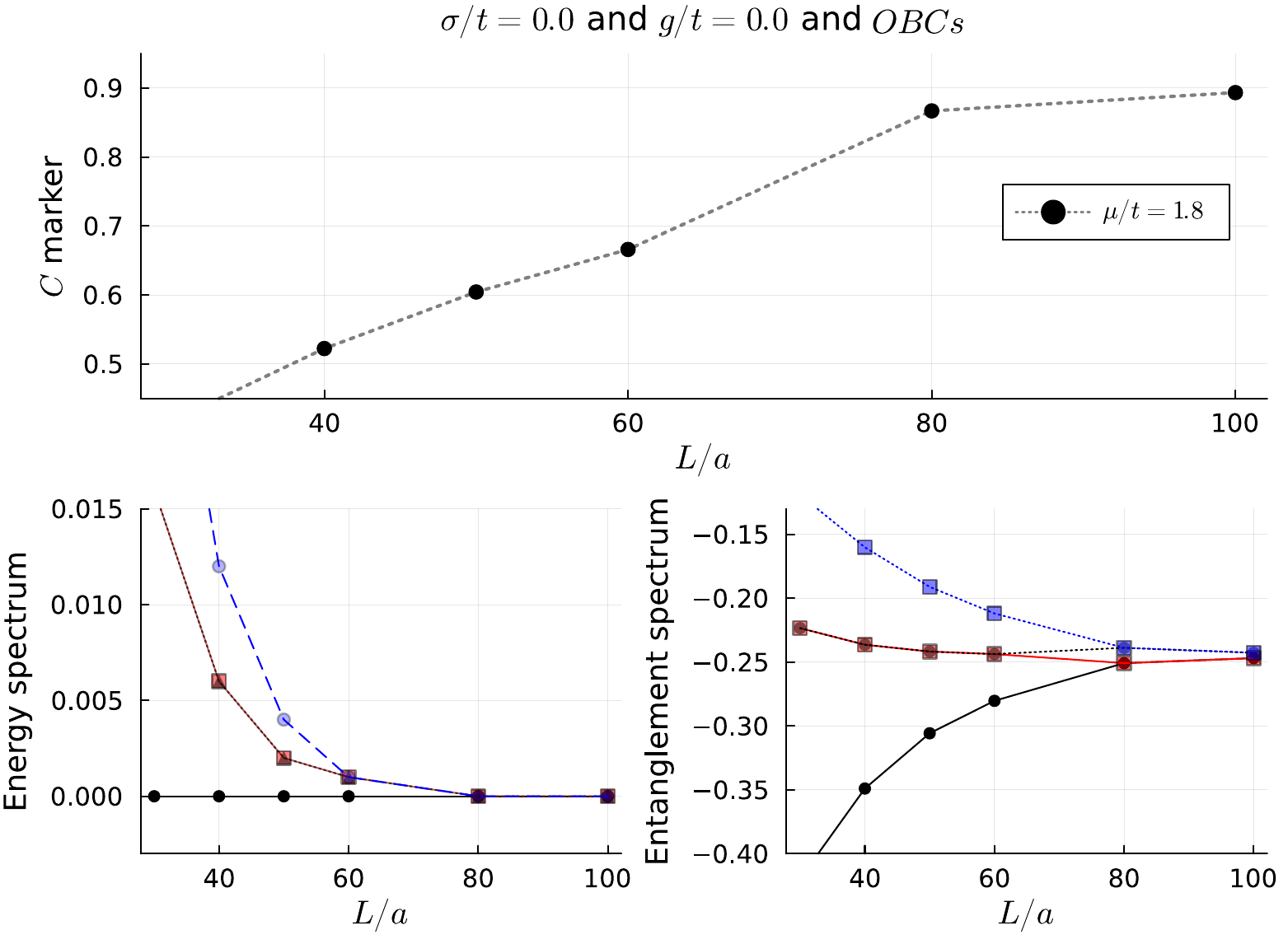}
	\caption{Clean and non-interacting case at $\mu/t = 1.8$ for $t = \Delta =1.0$. Edge mode overlap results in exponential splitting of the GS spectrum. Whilst the energy spectrum becomes four-fold degenerate already at $N = 80$ sites, the entanglement spectrum remains doubly degenerate. Increasing $N$ further we see how $C$ remains sensitive to the entanglement spectrum, rather than the energy gap.}
	\label{fig:ChernOBCquant}
\end{figure}

Firstly, for the clean and non-interacting wire, we note that the deviation from a quantized value $C = 1$ seems to be primarily due to the finite splitting of the entanglement spectra, as opposed to the GS energy. As can be seen in figure \ref{fig:ChernOBCquant} below, for a comparison of different system sizes at $t = \Delta$ and $\mu/t = 1.8$ (ie. close to the QCP), the Chern marker still improves toward $C = 1$ for $N=100$ sites, despite the GS spectrum being fourfold degenerate within presented accuracy. A comparison to the entanglement spectrum reveals a finite splitting, with the spectrum only being two-fold degenerate as opposed to the expected four-fold. This suggests the Chern marker acts also as an ``entanglement'' marker, which is not counterintuitive given the fact that all orders of long-range correlations contribute equally to $C$. 
\begin{figure}[h!]
	\centering
	\includegraphics[width = \linewidth]{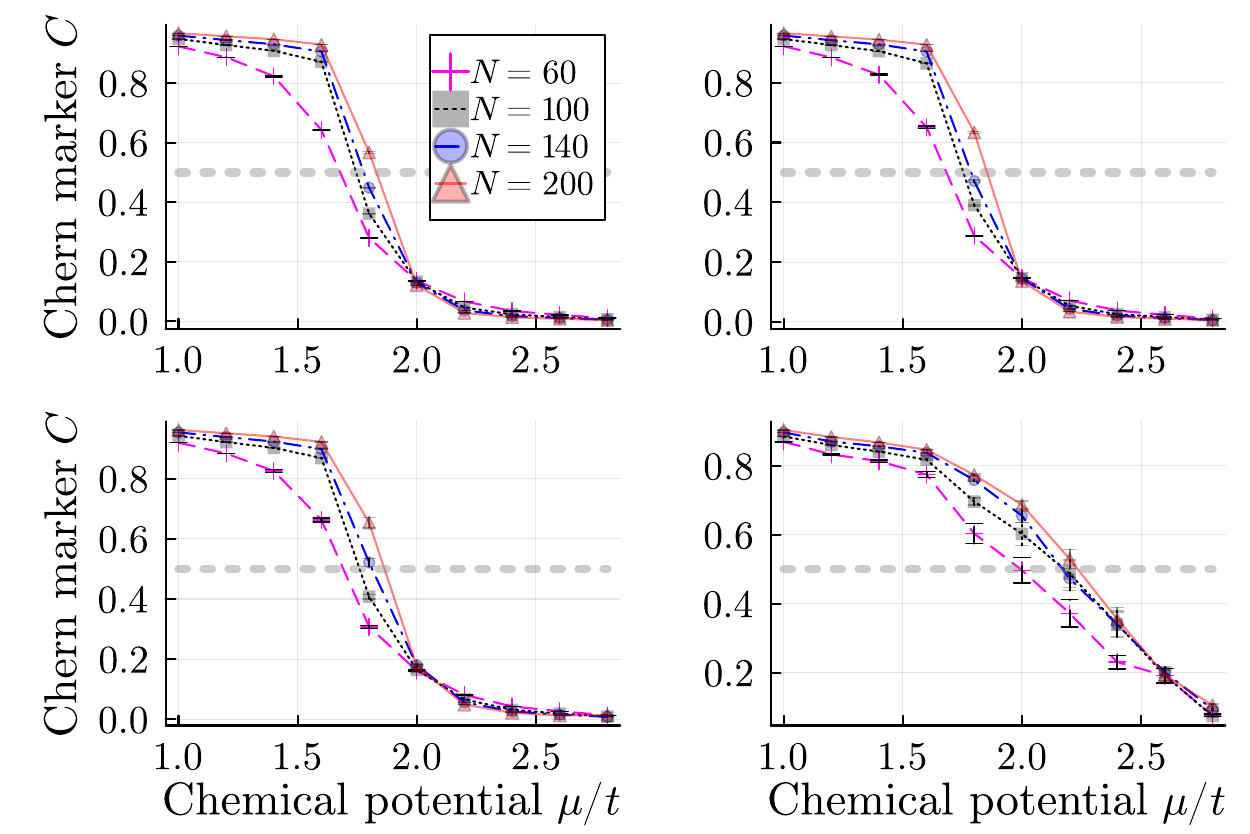}
\caption{Interaction $g/t = 2.0$ and $t = \Delta = 1.0$. Averaged $C$ marker for $M = 50$ realisations compared accross different system sizes.}
	\label{fig:avC2}
\end{figure}

Whilst the Chern marker improves drastically compared to smaller values of $N$, we find in figure \ref{fig:avC2} that for a wide range of disorder strengths system sizes of $N \geq 100$ already provide a very robust and precise phase boundary. Right at the phase transition and for large values of $\sigma/t$, a slight improvement can be observed up to $N = 200$ sites per wire, however overall the averaged Chern marker already captures the critical chemical potential at the phase transition for moderate system sizes. However, for a sharp transition from $\langle C\rangle = 1 \longrightarrow 0$, chains with $N\geq 200$ are necessary.
In figures \ref{fig:Delta_Comp} and \ref{fig:OBCvsDis_comp2} we show a comparison between $\Delta/t = 1.0$ and $\Delta/t = 0.3$ for $g/t = 2.0$ and $\sigma/t =0.3$. As can be seen, the $\Delta/t = 2.0$ case has less pronounced fluctuations, which supports the analytical (RG) results in \ref{AppendixB1} --- a larger effective SC gap is more robust against moderate disorder. 
\newpage

{\color{blue}
\subsection{Additional figures}}

\begin{figure}[h!]
	\centering
	\includegraphics[width = \linewidth]{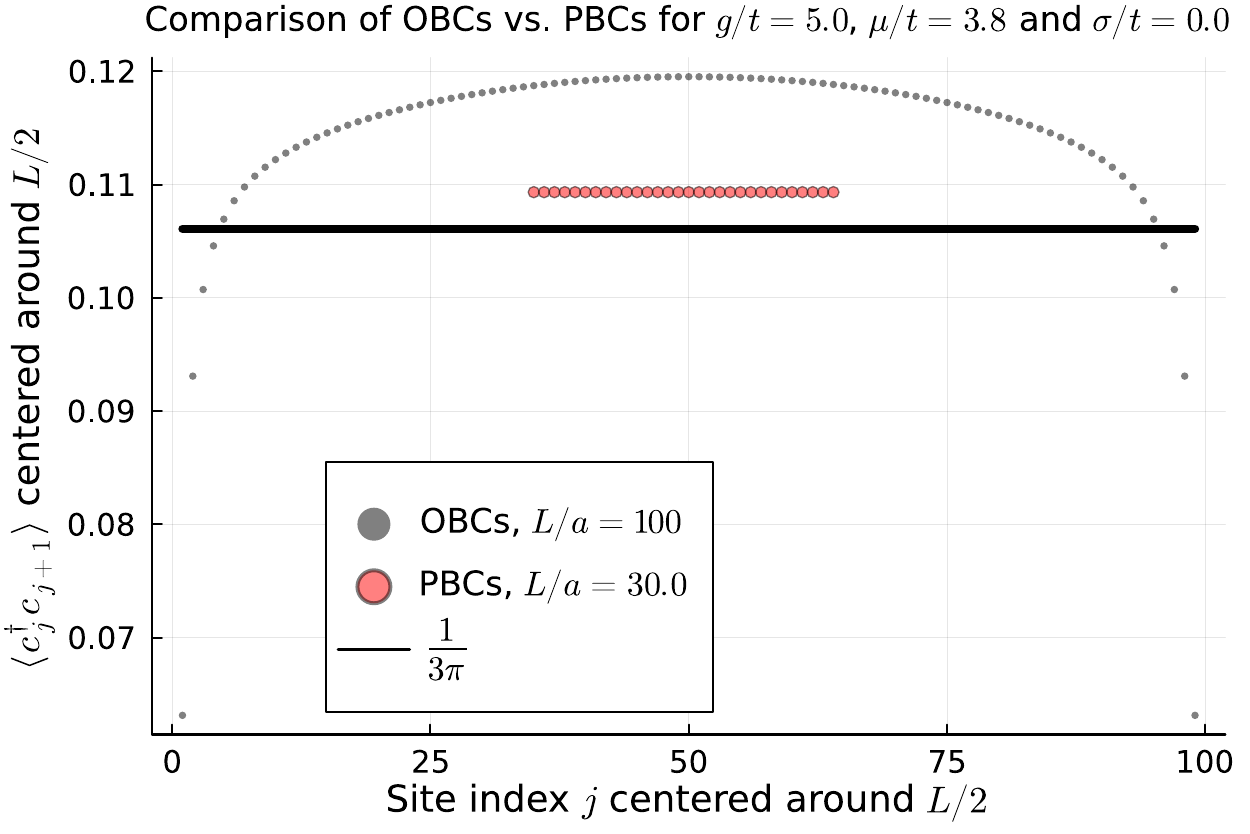}
	\caption{ Correlation functions $\langle c^{\dagger}_{j+1}c_{j}\rangle$ in the clean case ($\sigma/t = 0.0$) on wire $1$ for $g/t = 5.0$ at $\mu/t = 3.8$ with $t = \Delta = 1.0$. Results for PBCs were obtained for $N = 30$ sites per wire, whilst OBCs for $N = 100$. The solid black line is the analytical result for PBCs $\frac{1}{3\pi}$ at the QCP. \freplace{While local observables do not characterize the gapless critical phase, the strong correspondence between the non-interacting critical values and at large $g/t$ reinforces the view that the DCI phase is indeed simply an extension of the non-interacting QCPs, with similar local observables (Appendix \ref{AppendixA}).}}
	\label{fig:champagne_plots_PBC}
\end{figure}

\begin{figure}[h!]
	\centering
	\includegraphics[width = 1.0\linewidth]{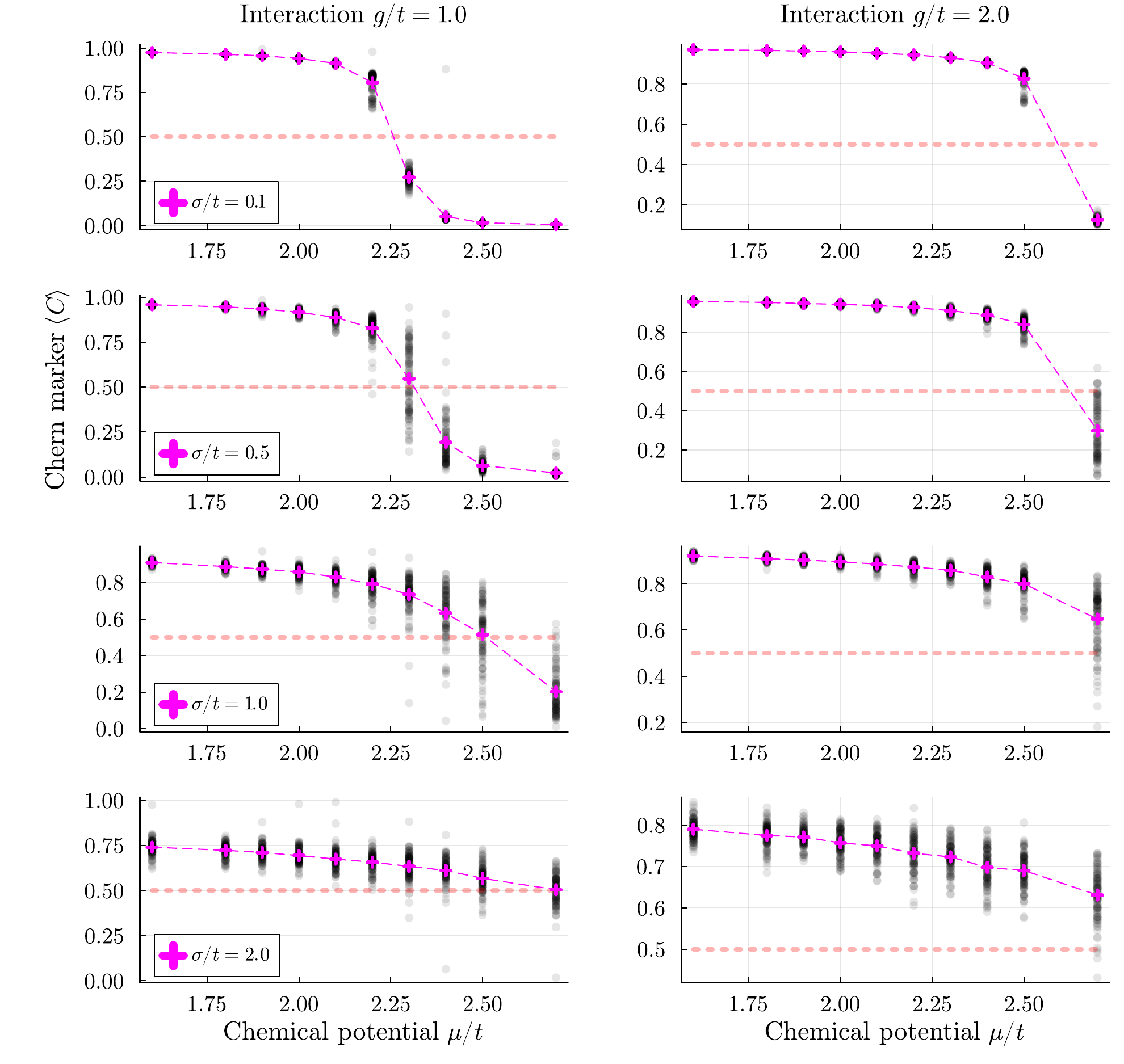}
\caption{Averaged $C$ marker for $M = 100$ realisations and $N = 140$ sites. (Black) Individual data points, (pink) the averaged $C$ value and (red) the $\langle C \rangle = 0.5$ critical value. }
	\label{fig:Delta_Comp}
\end{figure}
\begin{figure}[h!]
	\centering
	\includegraphics[width = \linewidth]{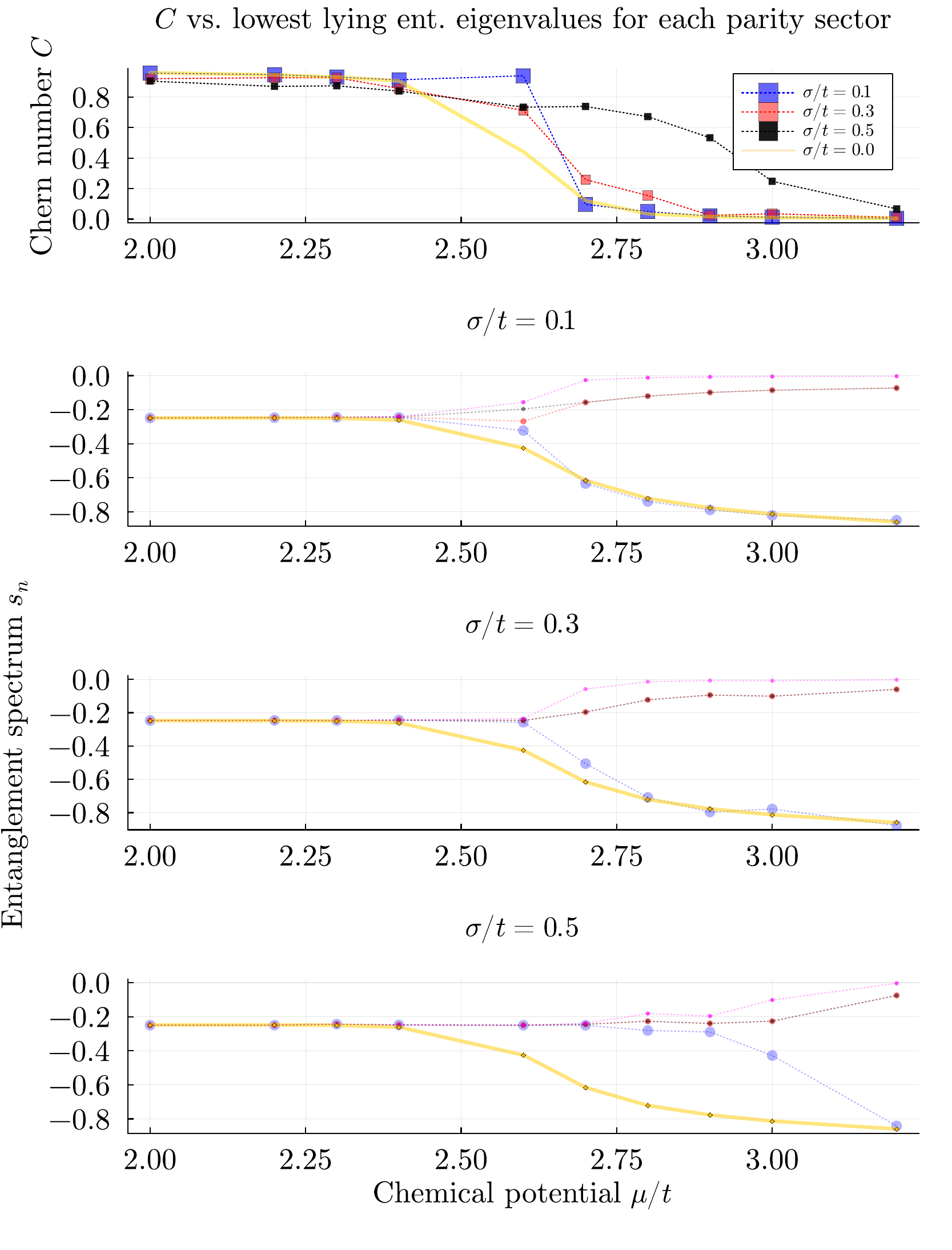}
\caption{{\color{black}Single realisations of disorder for $N = 140$, $t = \Delta$ and $g/t = 2.0$. The yellow curves show the case for $\sigma/t = 0.0$. 
The panels show: (Upper) Chern marker $C$, as a function of $\mu/t$ for a single realisation of disorder. (Lower) Lowest-lying entanglement eigenvalues for each parity-sector, and lying value for clean case in yellow.}} 
	\label{fig:OBC_Comp_N140}
\end{figure}

\begin{figure}[h!]
	\centering
	\includegraphics[width = \linewidth]{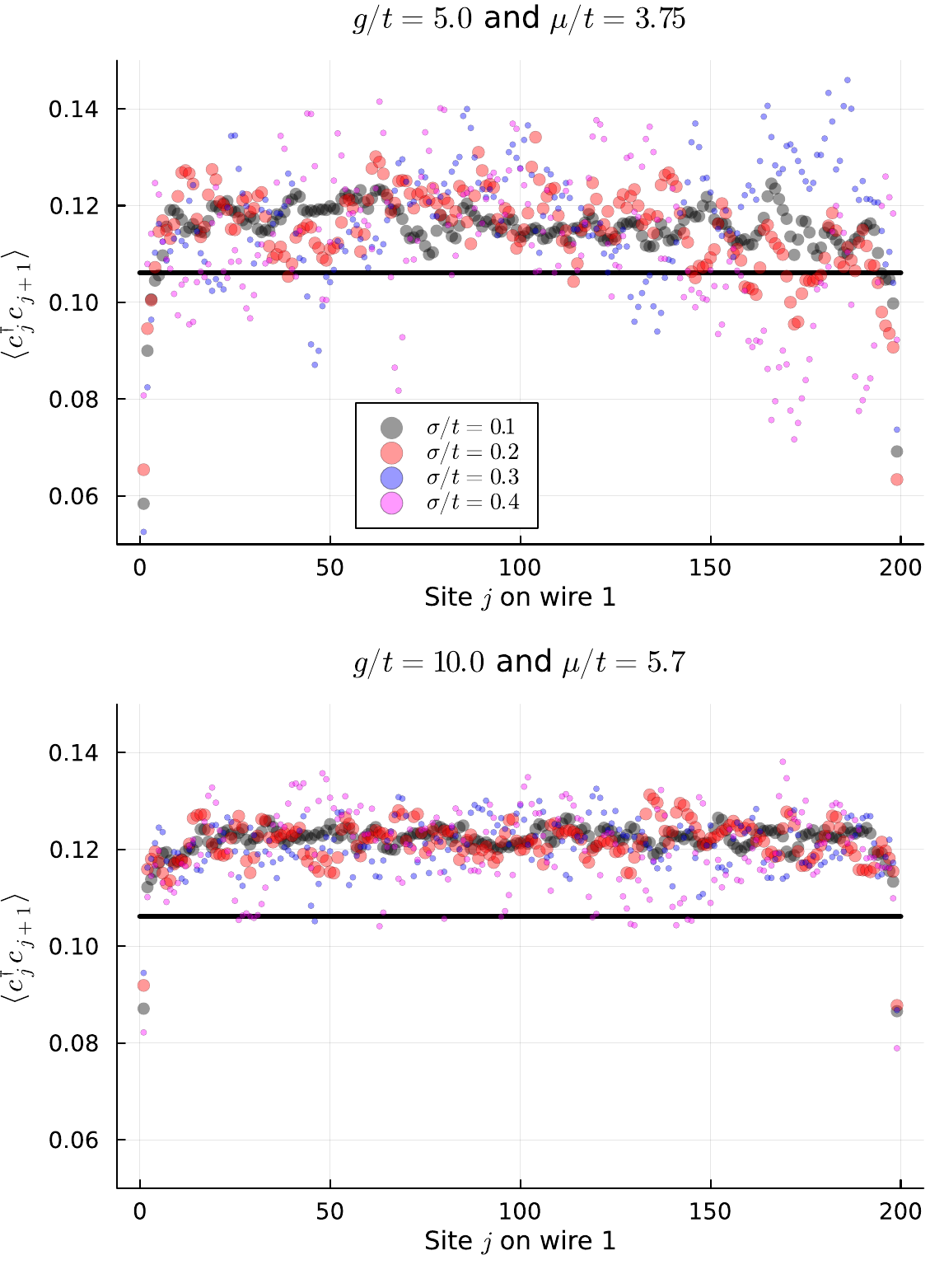}
	\caption{Correlation functions $\langle c^{\dagger}_{j+1}c_{j}\rangle$ on wire $1$ for $g/t = 5.0$ and $g = 10.0$ and $N = 200$ with OBCs and $t = \Delta$. Individual disorder realisations at $\mu/t = 3.75$ and $\mu/t = 5.7$. OBCs give a slightly different value compared to the predicted (PBC) value of $\frac{1}{3\pi}$ (black line), \emph{cf.} Appendix \ref{AppendixC}.}
	\label{fig:champagne_plots}
\end{figure}

\begin{figure}[h!]
	\centering
	\includegraphics[width = 0.75\linewidth]{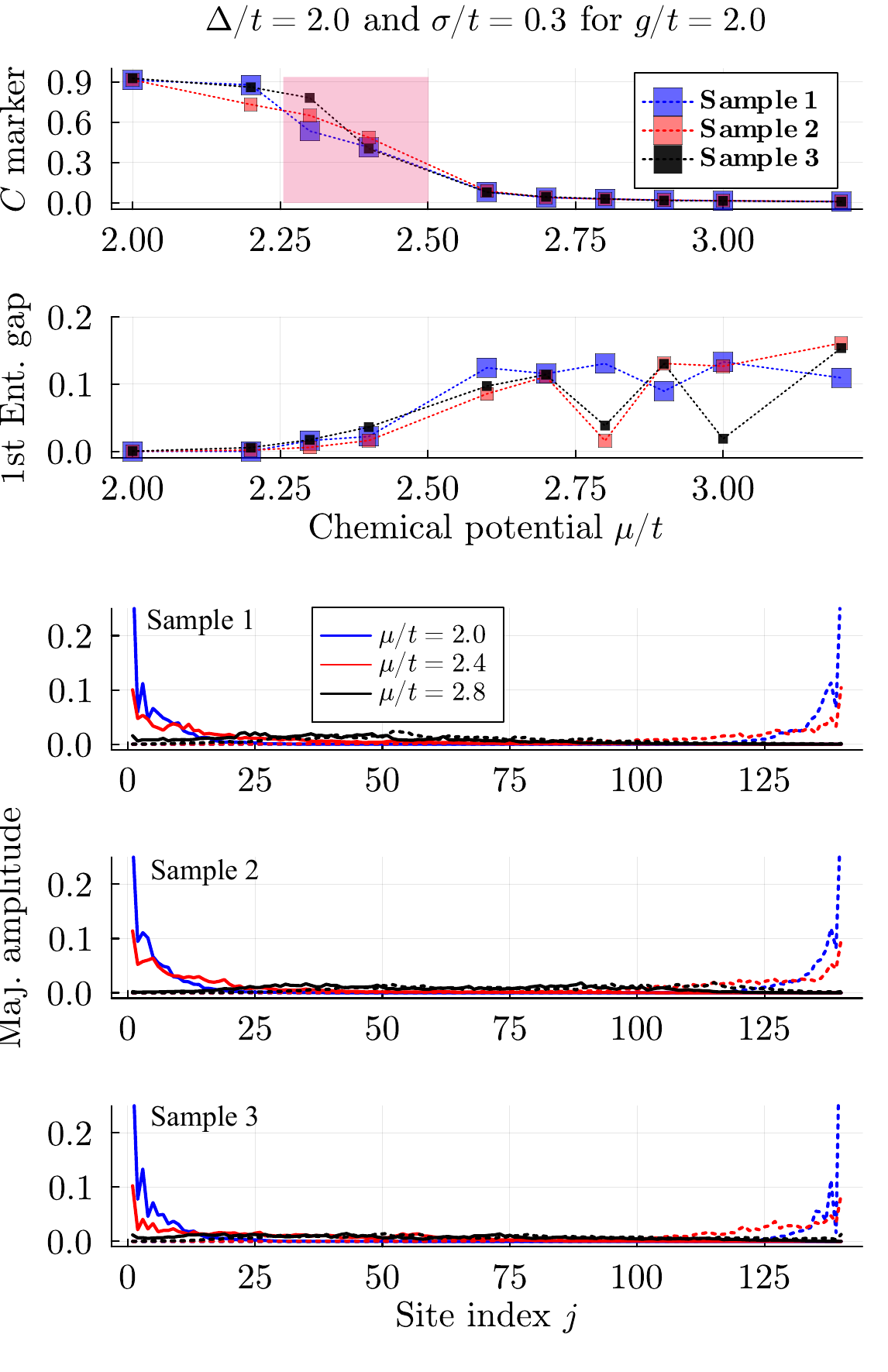}
\caption{Single realisations of disorder for $N = 140$, $\Delta = 2t$ and $g/t = 2.0$ and $\sigma/t = 0.3$. The shaded (pink) region in the first figure shows again the ``fuzzy" region in \ref{fig:PhaseDiagramlowdis}.
}
	\label{fig:OBCvsDis_comp2}
\end{figure}
\begin{figure}[h!]
	\centering
	\includegraphics[width = 0.75\linewidth]{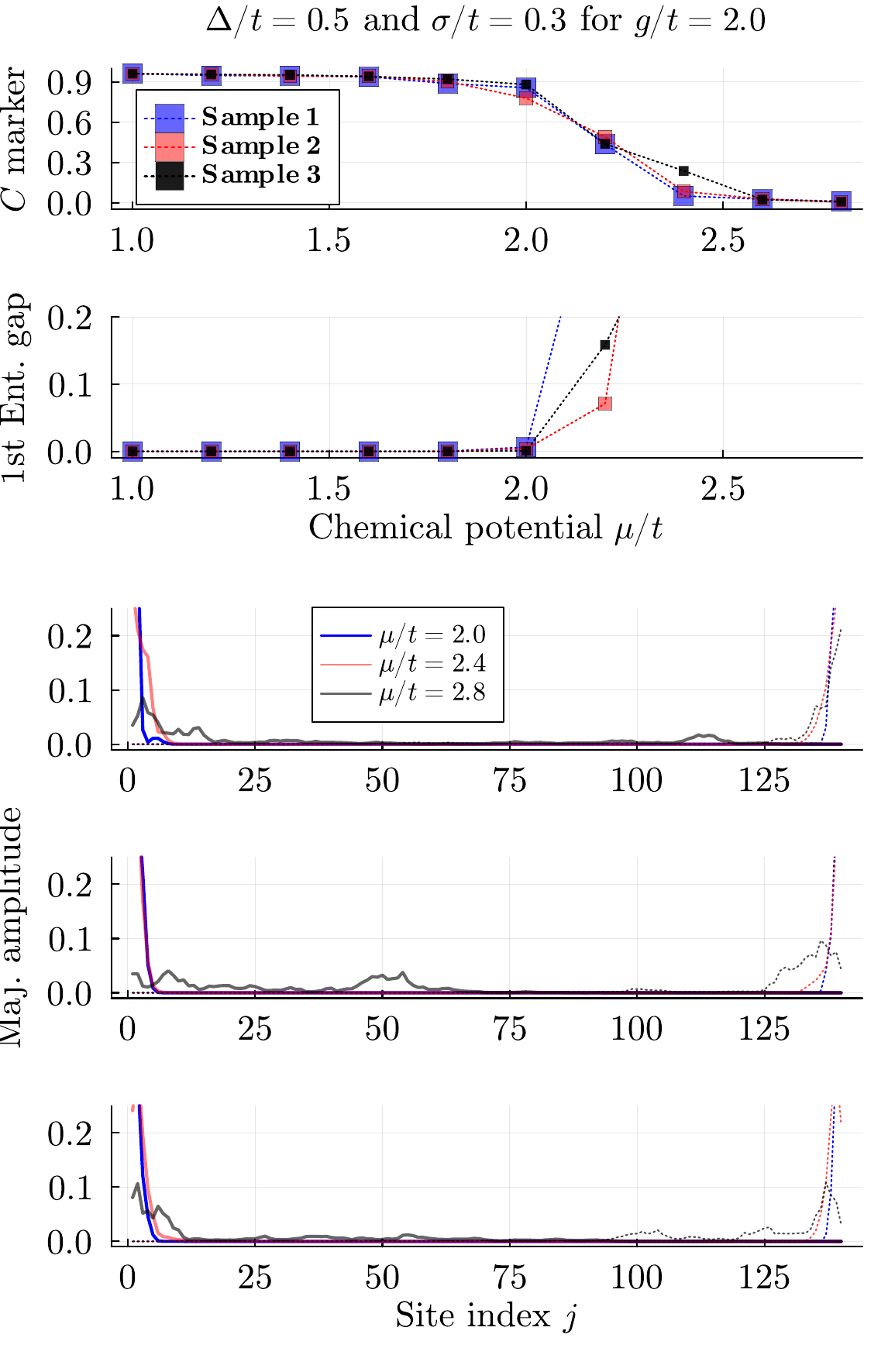}
\caption{Single realisations of disorder for $N = 140$, $\Delta = 0.5t$ and $g/t = 2.0$ and $\sigma/t = 0.3$.
}
	\label{fig:OBCvsDis_comp2}
\end{figure}
\begin{figure}[h!]
	\centering
	\includegraphics[width = 0.75\linewidth]{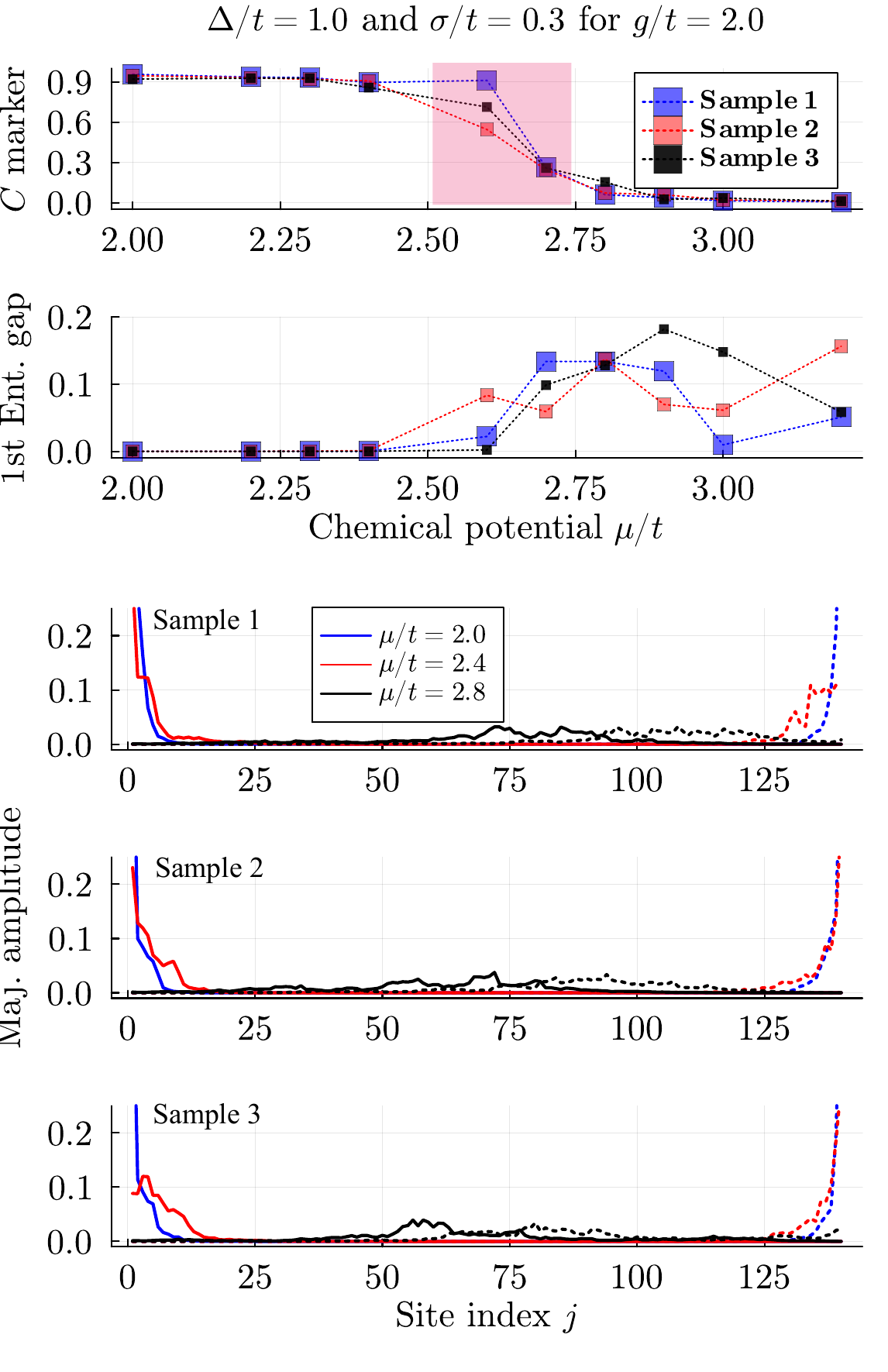}
\caption{Single realisations of disorder for $N = 140$, $t = \Delta$ and $g/t = 2.0$ and $\sigma/t = 0.3$. The shaded (pink) region in the first figure shows the ``fuzzy" region in \ref{fig:PhaseDiagramlowdis}. Outside of this region, the Chern marker, the first entanglement gap and the Majorana edge mode amplitudes clearly coincide. In the fuzzy region, we observe strong fluctuations of the $C$ marker and washed out Majorana amplitudes. In this region, finite-size effects and Majorana overlaps play a crucial role.}
	\label{fig:OBCvsDis_comp}
\end{figure}

\clearpage
\bibliography{OneWireDisorder}

\end{document}